\documentclass[twoside,english]{elsarticle}
\usepackage{ae,aecompl}
\usepackage[T1]{fontenc}
\usepackage[latin9]{inputenc}
\usepackage{geometry}
\usepackage{array}
\usepackage{float}
\usepackage{units}
\usepackage{multirow}
\usepackage{amsthm}
\usepackage{amsmath}
\usepackage{amssymb}
\usepackage{graphicx}

\makeatletter

\providecommand{\tabularnewline}{\\}

\theoremstyle{plain}
\newtheorem{thm}{\protect\theoremname}
\theoremstyle{definition}
\newtheorem{defn}[thm]{\protect\definitionname}
\theoremstyle{definition}
\newtheorem{example}[thm]{\protect\examplename}

\makeatletter
\def\ps@pprintTitle{%
  \let\@oddhead\@empty
  \let\@evenhead\@empty
  \def\@oddfoot{}
  \let\@evenfoot\@oddfoot
}
\makeatother

\usepackage{fancyhdr}
\usepackage{lineno}

\usepackage{pdflscape}
\usepackage[nolist,nohyperlinks]{acronym}

\makeatother

\usepackage{babel}
\providecommand{\definitionname}{Definition}
\providecommand{\examplename}{Example}
\providecommand{\theoremname}{Theorem}

\begin{document}
\begin{frontmatter}

\title{{\Large{}Aggregation operators for the measurement of systemic risk}\tnoteref{mytitlenote}}

\tnotetext[mytitlenote]{The paper is complemented with a web-based application: http://vis.risklab.fi/\#/fuzzyAgg.
The authors thank Gregor von Schweinitz and Tuomas Peltonen for comments
and discussions. The paper has also benefited from comments during
a presentation at IAMSR, Åbo Akademi University in Turku on November
13, 2014. Corresponding author: Peter Sarlin, Goethe University, Center
of Excellence SAFE, Grüneburgplatz 1, 60323 Frankfurt am Main, Germany.
E-mail: peter@risklab.fi.}

\author[H]{Jozsef Mezei}

\author[P1,P2,P3]{and Peter Sarlin}

\address[P1]{Center of Excellence SAFE at Goethe University Frankfurt, Germany}

\address[P2]{Department of Economics, Hanken School of Economics, Helsinki, Finland}

\address[P3]{RiskLab Finland at Arcada University of Applied Sciences, Helsinki,
Finland}

\address[H]{Institute for Advanced Management Systems Research (IAMSR), Åbo
Akademi University, Turku, Finland}
\begin{abstract}
The policy objective of safeguarding financial stability has stimulated
a wave of research on systemic risk analytics, yet it still faces
challenges in measurability. This paper models systemic risk by tapping
into expert knowledge of financial supervisors. We decompose systemic
risk into a number of interconnected segments, for which the level
of vulnerability is measured. The system is modeled in the form of
a Fuzzy Cognitive Map (FCM), in which nodes represent vulnerability
in segments and links their interconnectedness. A main problem tackled
in this paper is the aggregation of values in different interrelated
nodes of the network to obtain an estimate systemic risk. To this
end, the Choquet integral is employed for aggregating expert evaluations
of measures, as it allows for the integration of interrelations among
factors in the aggregation process. The approach is illustrated through
two applications in a European setting. First, we provide an estimation
of systemic risk with a of pan-European set-up. Second, we estimate
country-level risks, allowing for a more granular decomposition. This
sets a starting point for the use of the rich, oftentimes tacit, knowledge
in policy organizations.\end{abstract}
\end{frontmatter}

\newpage{}

\textit{\footnotesize{}\hspace{3cm}\textquotedblleft }\emph{\footnotesize{}Clearly,
there is widespread awareness that the analytical framework, no matter
its level of }{\footnotesize \par}

\textit{\footnotesize{}\hspace{3cm}}\emph{\footnotesize{}sophistication,
cannot replace the expert knowledge and judgment of the ESRB\textquoteright s
members.}\textit{\footnotesize{}\textquotedblright{}}{\footnotesize \par}

\textit{\footnotesize{}\hspace{3cm}}{\footnotesize{}-- Vítor Constâncio,
Vice-President of the ECB, Frankfurt am Main, 16/11/2010}{\footnotesize \par}

\section{Introduction}

Measurement of systemic risk has become a pivotal topic among academics,
policymakers and supervisors. The search for the one unrivaled systemic
risk measure has mostly stimulated empirical research for a mechanistic
analysis of system-wide risks. Exploiting the fact that macroprudential
supervisory authorities possess a variety of specialized domain intelligence
and experience, this paper takes a bottom-up approach to address the
topic: \emph{How do we tap into the expertise of individual supervisors
to measure systemic risk?}

The current financial crisis has highlighted the importance of a macroprudential
approach to ensuring financial stability \citep{Borio2011}. In contrast
to only being concerned with the stability of individual financial
institutions (i.e., microprudential), the shift towards a system-wide
perspective has imposed complexity in terms of analysis tasks and
the underlying data (see \citet{FloodMendelowitz2009}). It accentuates
the need for an understanding of not only individual financial components,
be they economies, markets or institutions, but also interconnectedness
among them and their system-wide risk contributions. To this end,
analytical tools and models provide ample means for two types of tasks:
(\emph{i}) early identification of vulnerabilities and risks, as well
as their triggers, across financial instruments, markets and institutions,
and (\emph{ii}) early assessment of transmission channels of and a
system's resilience to shocks, and potential severity of the risk
materialization. Yet, despite the rise of big data and analytics in
previous years, macroprudential analysis as a support to policy remains
highly dependent upon market intelligence and expert judgment and
experience, as is above noted by Vítor Constâncio. An illustrative
and intuitive example is the ever increasing shadow banking activities
occurring behind the scenes, in which quantitative risk analysis and
measurement are challenging tasks. Going beyond lack of data, one
could in line with Lucas' critique and Goodhart's law also question
the use of quantitative models in an ever changing environment, such
as the impact of regulation on markets and the endogeneity of risk
(e.g., \citet{DanielssonShin2003}).

Managing knowledge within an organization in an efficient way is an
essential capability, not the least for knowledge-producing organizations
like macroprudential supervisory bodies. In addition to producing
information about systemic risks, a key task is to disseminate it
horizontally and vertically within the organization. Yet, expert knowledge
is not unproblematic. Beyond common challenges in the incentive structure
to share information (e.g., \citet{Lin2007}), a large share of challenges
relate to the fact that most knowledge possessed by experts is classified
as unstructured and tacit (e.g., \citet{Haldin2000}). This obviously
hinders capturing, representing, and transferring it within the organization.
Leveraging on groups of experts knowledgeable in specific topics,
a key concern ought to be judging (\emph{i}) which expert\textquoteright s
knowledge is more relevant or reliable than the others, and (\emph{ii})
how to combine the knowledge of different experts in a structured
way to obtain a unique solution to a problem. One solution to these
types of challenges comes from the family of aggregation operators.
To this end, we need to answer the remaining question: \emph{How do
we aggregate expert opinions to measure systemic risk?}

The quantification of expert knowledge in risk assessment is not uncommon
(e.g., \citep{Goossens2001}). In this paper, the objective is to
provide a framework for measuring systemic risk by aggregating the
knowledge of financial supervisors with a set of families of aggregation
operators. Since its introduction, fuzzy set theory and fuzzy logic
has been applied in numerous contexts and in numerous ways to utilizing
expert knowledge. The most traditional way involves creating fuzzy
rule-based systems based on the knowledge extracted from the experts.
An example of this type of research can be found in \citet{Leon2013}
and \citet{Leon2013a}, whose focus is closely connected to the focus
of this paper. The authors propose to use a fuzzy inference system
with linguistic rules to estimate the systemic importance of financial
institutions; the rules are extracted from expert\textquoteright s
assessment of the possible combinations of factors describing the
system and combined using different approximate reasoning schemes.
A different direction of applications makes use of fuzzy sets and
fuzzy measures in representing and aggregating expert knowledge with
the main problem being the appropriate choice of aggregation function,
as highlighted in \citet{Beliakov2001}. As is pointed out by \citet{Moon1999},
the two main issues to consider in these approaches are to decide
(\emph{i}) in what form the information is elicited from the experts,
and (\emph{ii}) how the information is aggregated in the presence
of multiple experts. 

In this paper, we present an approach that combines Fuzzy Cognitive
Maps (FCMs) and aggregation based on Choquet integrals to handle the
two above mentioned challenges. The first important issue to tackle
as the basis of the aggregation process is to identify a representation
of data acquired from experts (i.e., the output of the knowledge elicitation
process). The representation of the expert knowledge should not only
be such that it provides an appropriate placeholder for the effective
use of the aggregation process, but it should also support analyzing
the system underlying the problem from different perspectives. The
general goal is to estimate the overall level of risk present in a
complex system, in our case the vulnerability of the financial system.
The system can be described as the hierarchy of components. A network
representation of the system, with the components of the system as
the nodes and the interrelations between the components as the edges,
provides flexibility to analyze different attributes of the system,
such as identifying the most critical or central nodes and their interlinkages.
In this paper, the FCM \citep{Kosko1986}, as a special type of a
weighted graph, is utilized to capture and make use of expert evaluations
regarding the interrelation between different sectors of a financial
system. As our main goal is to estimate systemic risk and not, as
in general applications of FCMs, to identify the ``optimal'' input
vector of the map that results in an equilibrium point, we focus on
aggregating the impacts of a set of nodes on a target node of the
network. For this purpose, we need to identify an appropriate aggregation
operator that can handle the complex interrelations in the underlying
system and provide a final value of systemic risk. In this paper,
we show how we can model the spread of risk in the system, represented
as a FCM, through a fuzzy measure and consequently use the corresponding
Choquet integral to aggregate values in the nodes of the map. With
our approach, we not only provide a measure of systemic risk but also
identify the most central parts of a system, such as the most vulnerable
components or countries. We achieve this goal by specifying different
fuzzy measures corresponding to different components of the modeled
system. The main theoretical contribution of the article lies in combining
FCMs and Choquet integrals to represent and analyze complex systems
of interrelated objects. Additionally, we propose approaches for assessing
the system through quantitative network measures and visual network
graphs.

We illustrate the approach for measuring systemic risk from expert
opinions through two applications in a European setting, for which
we also discuss practical implications and challenges. First, we provide
an estimation of systemic risk in a pan-European set-up, where we
model systemic risk at the European, country and sectoral level. Second,
we also estimate country-level risk, allowing for a more granular
decomposition, by modeling risk at the level of the country, its sectors
and sub-dimensions of the sectors. For both applications, we also
perform quantitative and qualitative analysis of the systemic risk
measures as a network of nodes and edges. While the former uses standard
measures of network centrality, the latter approach provides visual
interactive interfaces for the systemic risk measures. The visualizations
are available as a web-based application.%
\footnote{The complimentary web-based applications are available here: http://vis.risklab.fi/\#/fuzzyAgg%
} The approach overall and these applications in particular set a starting
point for the use of the rich, oftentimes tacit, knowledge in a policy
setting overall and their organizations in particular.

The rest of the paper is structured as follows. Section 2 links systemic
risk to expert knowledge of supervisors and policymakers, as well
as introduces aggregation operators. In Section 3, we introduce FCMs
and the use of the Choquet integral, particularly for the measurement
of systemic risk. Section 4 presents two applications: the case of
systemic risk in Europe and in an individual country. Finally, we
conclude and discuss future research in Section 5.

\section{Systemic risk and aggregation operators}

This section discusses the concept of systemic risk, and the use of
aggregation operators for the task of its measurement. After providing
an overview of systemic risk, we provide a brief overview of aggregation
operators, as well as a mapping back into the task of systemic risk
measurement.

\subsection{Macroprudential supervision and systemic risk}

A key task of a macroprudential supervisory body is to internally
collect and produce information about systemic risk. The term systemic
risk, not to paraphrase Justice Potter Stewart's definition of explicit
content \citep{Bisiasetal2012}, belongs to the group of concepts
that are broad and vague, yet implicitly understood. Still, we need
a working definition as a basis for measurement and analysis. To start
with, financial instability is defined as an event that has adverse
effects on a number of important financial institutions or markets
(ECB \citep{ECB2009}). Systemic risk, as also defined by the ECB,
is the risk of widespread financial instability that impairs the functioning
of the financial system to the extent that it has severe implications
on economic growth and welfare.

\begin{figure}[H]
\begin{centering}
\includegraphics[width=0.8\columnwidth]{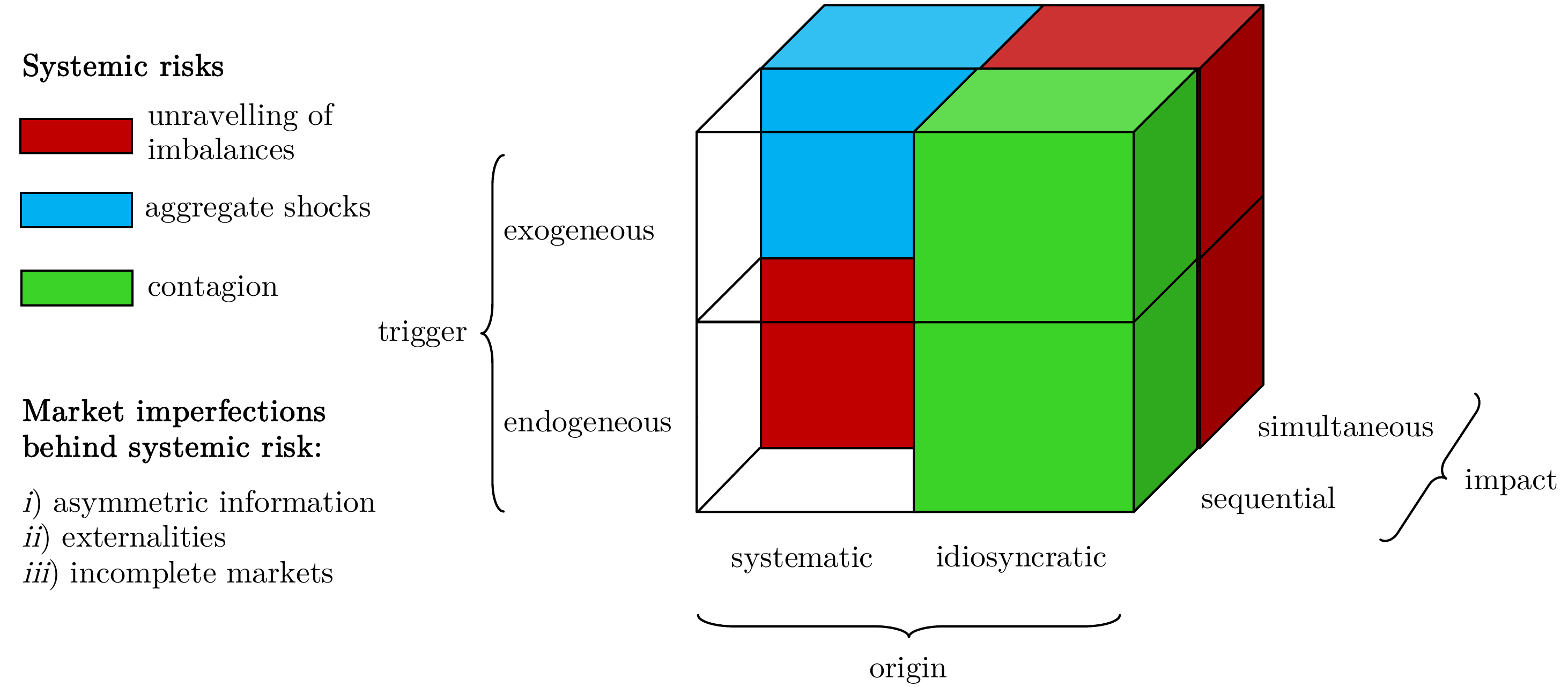}
\par\end{centering}

\textbf{\scriptsize{}Notes}{\scriptsize{}: The figure represents the
systemic risk cube with three dimensions and systemic risks and is
an adapted version of that in \citet{ECB2010}.}{\scriptsize \par}

\centering{}\protect\caption{\label{Fig:sys_risk}Systemic risk cube with three forms of risks.}
\end{figure}

The definition used herein is untangled with the help of the systemic
risk cube shown in Figure \ref{Fig:sys_risk}. The notion of a risk
cube was introduced by the \citet{ECB2010}, and represents their
conceptual framework, but has its origin in a number of works.%
\footnote{For further information, see \citet{deBandtHartmann2002}, \citet{deBandt2009},
\citet{ECB2009}, and \citet{ECB2010}%
} The three dimensions of the risk cube are the triggers, origins and
impacts. The nature of \emph{triggers} unleashing the crisis could
take the form of an exogenous shock, which stems from the outside
of the financial system (e.g., macro-economic shocks) or could emerge
endogenously from within the financial system (e.g., banks). The \emph{origins}
of the events may be distinguished to limited idiosyncratic shocks
and widespread systematic shocks. While the former initially affect
only the health of a single financial market, financial intermediary
or asset, the latter may in the extreme affect the entire financial
system. Further, the \emph{impact} of the events may cause problems
for a range of financial intermediaries and markets in a sequential
and simultaneous fashion.

Beyond three dimensions, we herein concretize the notion of systemic
risk through the three forms presented by \citet{deBandt2009}. The
first form of systemic risk focuses on the unraveling of \emph{widespread
imbalances} in the vein of Kindleberger\textquoteright s \citep{Kindleberger1996}
and Minsky\textquoteright s \citep{Minsky1982} financial fragility
view of a boom-bust credit or asset cycle. Hence, the subsequent abrupt
unraveling of the imbalances may be endogenously or exogenously caused
by idiosyncratic or systematic shocks, and may have adverse effects
on a wide range of financial intermediaries and markets in a simultaneous
fashion. Early and later empirical literature alike have identified
common patterns in underlying vulnerabilities preceding financial
crises (e.g., \citet{Kaminskyetal1998b} and \citet{ReinhartRogoff2008}).
The second type of systemic risk refers to a widespread \emph{exogenous
aggregate shock} with negative systematic effects on one or many financial
intermediaries and markets at the same time. These types of aggregate
shocks have empirically been shown to co-occur with financial instabilities
(e.g., \citet{Gorton1988} and \citet{DemirgucDetragiache1998}),
and can be exemplified by the collapse of banks during recessions
due to the vulnerability to economic downturns. The third form of
systemic risk is \emph{contagion and spillove}r, which usually refers
to an idiosyncratic problem, be it endogenous or exogenous, that spreads
in a sequential fashion in the cross section. There is wide evidence
of cross-sectional transmission of financial instability (e.g., \citet{UpperWorms2004}
and \citet{LelyveldLiedorp2006}), such as the failure of one financial
intermediary causing the failure of another, which initially seemed
solvent, was not vulnerable to the same risks and was not subject
to the same original shock as the former. While contagion refers to
a situation when the initial failure is entirely responsible for subsequent
ones, the term spillover is commonly used when the causal relationship
is unknown (e.g., \citet{ECB2010}).

Macroprudential oversight requires a broad toolbox of models for systemic
risk measurement. The categorization by \citet{ECB2010} elegantly
maps the three forms of systemic risk to analytical tools: (\emph{i})
early-warning models, (\emph{ii}) macro stress-testing models, and
(\emph{iii}) contagion and spillover models. First, to identify vulnerabilities
and imbalances in an economy, \emph{early-warning models} derive probabilities
of the future occurrence of systemic financial crises (e.g., \citet{Alessi2011520}
and \citet{Duca2012}). With a set of vulnerability indicators as
an input, the output of such models takes the form of crisis probabilities,
which are monitored with respect to critical threshold values. Second,
\emph{macro stress-testing models} provide means to assess the resilience
of the financial system to a variety of aggregate shocks (e.g., \citet{Castrenetal2009}
and \citet{Hirtleetal2009}). These exercises assess the consequences
of assumed extreme, but plausible, shocks for different entities,
for which a key question is to find the balance between plausibility
and severity of the stress scenarios (e.g., \citet{AlfaroDrehmann2009}
and \citet{Quagliariello2009}). Third, \emph{contagion and spillover
models} can be employed to assess how resilient the financial system
is to cross-sectional transmission of financial instability (e.g.,
\citet{IMF}). Hence, they attempt to answer the question: With what
likelihood, and to what extent, could the failure of one or multiple
financial intermediaries cause the failure of other intermediaries?
Beyond measures of systemic risk, coincident indicators measure the
contemporaneous level of systemic stress (e.g., \citet{Holloetal2012}),
which may be used to identify, signal and report on heightened stress.

The three types of systemic risk, and the corresponding tools for
measuring them, provide a starting point for an all-encompassing framework
of systemic risk. At a more granular level than the forms of systemic
risk, measurement ought to be broken down to specific market segments
or economic sectors and countries or other geographical definitions.
Hence, for each market segment and economy, and at each point in time,
the following characteristics should be measured: (\emph{i}) specific
imbalances building-up in the cross-section and their current state,
the likelihood of these imbalances to unravel, and their potential
severity; (\emph{ii}) transmission channels of aggregate shocks, an
overview of plausible shocks, impacts on other market segments, and
potential severity of and resilience to shocks in case of materialization;
and (\emph{iii}) sources of contagion or spillover at the individual
and system-level, as well as potential severity of and resilience
to cross-sectional transmission. Mapping these to Borio's \citep{Borio2009}
two dimensions of systemic risk, the former relates to time and cyclicality,
as risk builds-up in tranquil times and abruptly unravels in times
of crisis, whereas the two latter relate to cross-sections and structures,
as risk may transmit through various channels in interconnected financial
systems. Accordingly, these two dimensions can also be mapped to the
two tasks of risk identification and assessment, and are what we ought
to be interested in when measuring risk with any approach.

\subsection{Systemic risk and expert knowledge}

Beyond only collecting and producing information about systemic risks,
a key internal task of a macroprudential supervisory body is to disseminate
it horizontally and vertically within the organization. Yet, to highlight
the importance of disseminating information vertically, most if not
all of the produced information has an external contact point only
through the very top of the organizational structure. The above discussed
analytical approaches, while giving an impression of purely mechanistic
analysis, is only a part of the truth. Independent of the rigor of
the analytical tools, expert knowledge is always a central part of
informed policy decisions. This is obviously due to reasons related
to transparency and accountability, as decision clearly are taken
by humans rather than models, but relates also to challenges faced
by policymakers in the measurement of some, if not all, types of systemic
risk.

At an epistemological level, one could even argue that predictions
about future human behavior is impossible, where one example is the
Lucas critique or Goodhart's law of regulation impacting behavior
and the endogeneity of risk another (e.g., \citet{DanielssonShin2003}).
Further, systemic risk is not restricted to banks \citep{Frenchetal2010}.
Since \citet{McCulley2007} coined the term shadow banking, it has
become widely recognized that credit intermediation outside the banking
system takes place in an environment where regulation and supervision
is applied to a lesser degree \citep{FSB2011}. Thus, the lack of
reporting standards and obligations for shadow banking activities
obviously directly impacts measurability. If data exist, a challenge
on its own is its quality, comparability and timeliness. Although
the US may have more standardized data, concerns have been arisen
about pan-European data comparability \citep{ESMA2013}, such as definitions
of non-performing loans and forbearance practices, not to mention
transatlantic comparability. Beyond precision, accounting-based data
also exhibits long publication lags.

When measurability is questioned, and even though it would not be,
an obvious qualitative resource for a policymaker would be individual
and organization-wide expert knowledge and judgment. Accordingly,
a report by \citet{IMFBISFSB2009} stresses that systemic risks cannot
be assessed with solely quantitative approaches, and that authorities
ought to draw upon intimate knowledge of their own financial system's
functioning. We define the term expert as someone who has long-term
experience via research and practice in a specific discipline or task.
Given the fact that central banks have a high knowledge concentration
due to well-educated and oftentimes narrowly focused work tasks, it
can also seen as an organization with a high share of expertise. In
addition to well-educated central bank governors, \citet{Fox2014}
highlights that out of 450 employees in research-oriented divisions
about half have a PhD in economics. In addition, central banks have
direct and explicit research functions, ideally with the ultimate
aim of supporting policy formulation, as discussed in \citet{Bercetal2009}.
Except for increases in the number of PhDs, and overall educational
standards, \citet{Fox2014} also highlights that the share of accountants
and lawyers has gradually been taken over by economists in the Federal
Open Market Committee. As \citet{Goodhartetal2002} show that central
banks employ more economists and fewer lawyers and accountants for
their own financial stability and supervisory functions than non-central
bank supervisory agencies do, this clearly correlates well with a
separation of monetary policy (macro) and bank regulation tasks (micro)
at the time. Given current macroprudential initiatives, we are likely
to see a shift back through an increase in the share of supervisory
tasks within or close to central bank activities. This accentuates
the heterogeneity in possessed qualitative knowledge that must be
managed at a macroprudential supervisory body.

The wealth of focused expertise, while in isolation only being beneficial
in individual in-depth investigations, would ideally be a valuable
resource as a support for the entire organization, as many tasks require
a multidisciplinary perspective. Within expertise-driven and knowledge-producing
organizations like macroprudential supervisory bodies, expertise obviously
ought to be disseminated both vertically and horizontally. To this
end, this paper taps into expert knowledge of systemic risk through
its aggregation, in order to foster and provide means to support knowledge
sharing and systemic risk identification and assessment overall.

\subsection{Aggregating expert knowledge}

Measuring or estimating systemic risk simply means forecasting the
likelihood of events that can affect the behavior of the entire system
under consideration. In many real life situations, statistical approaches
provide an important tool for assessing a system when one possesses
sufficient reliable and valuable historical data regarding also future
events. Yet, there are several cases when other types of methodologies
can be more beneficial. A typical choice of approach discussed extensively
for example by \citet{Cooke1991} is to rely on expert evaluations,
estimations or opinions. Expert opinions can be used most importantly
to assess the possibility of events with very low likelihood but high
potential impact. When the process involves multiple experts, rather
than opinion elicitation of one expert, the obtained opinions need
to be combined into a final assessment. There are several issues to
be considered regarding the aggregation procedure of individual expert
opinions. On the most general level, \citet{Clemen1999} classify
different approaches to two main groups: mathematical and behavioral.
In behavioral approaches, the experts are encouraged to interact among
each other with the final goal of obtaining a consensual solution
to the given problem (typical in group decision making). By mathematical
approaches in the context of risk estimation, the literature mainly
refers to the elicitation and aggregation of subjective probabilities.
This refers to any formal mathematical procedure for aggregation when
the experts\textquoteright{} input is restricted to evaluation without
further involvement in the process. In practice, the combination of
the two approaches are used most frequently.

From a mathematical perspective, the aforementioned view of considering
expert assessments as subjective probabilities already imposes assumptions
on the interpretation of an expert opinion as a type of likelihood.
Yet, when combined with rigorous mathematical procedures, they can
provide essential insights in many applications. Beyond this, there
can be different interpretations assigned to the expert opinions regarding
even when the nature of uncertainty they try to capture is the focus
of the aggregation process, as pointed out in \citet{Hoffman1994}.
For this reason, the use of traditional statistical methods for handling
expert opinions, such as Bayes modeling described by \citet{Winkler1968},
is not necessarily optimal or even justifiable. Likewise, \citet{Mosleh1988}
pointed out that in general relying only on the ``common sense''
of experts in a risk assessment process is not the best choice, but
their aggregation can provide essential insights in practice. In many
cases, the opinions of decision-makers can be interpreted in the framework
of preference and utility theories (\citet{Keeney1993d}). In these
cases, aggregation can be performed using traditional operators used
in decision-making problems (see \citet{Grabisch2009} for a comprehensive
treatise on the topics of aggregation functions). There are several
important factors related to the elicitation and aggregation process
that can be modeled by choosing an appropriate aggregation operator
(e.g., the difference in reliability or importance of experts).

Various domains and their specific problems require performing the
task of summarizing a set of numerical values differently. Obtaining
a single number that provides a representation of the original set
and corresponds to the predefined requirements is not trivial. A large
number of aggregation operators exist to summarize numerical and/or
non-numerical information into a single value. These operators play
a fundamental role in many fields related to different decision-making
problems. A typical application is the case when an evaluation has
to be performed in the presence of multiple objects to be assessed
individually and then the assessment is to be combined into an overall
evaluation. In many cases, independence of the object is assumed;
the most popular aggregation methods in these situations include a
wide range of operators from the simple weighted average to the OWA
(ordered weighted average) introduced by \citet{Yager1988} and used
in a wide range of real life problems. In these operators every evaluation
is considered individually in the aggregation process and performed
independently from others. Yet, the most general notion of an aggregation
operator requires only one property to be satisfied by the operator:
monotonicity. It implies that if a set of values to be aggregated
dominates a set of values, then the aggregated value for the dominated
set cannot be higher. In our context it means that if the risk level
of every component in a system increases, then the overall systemic
risk cannot decrease. Besides the assumption of monotonicity, every
other property can be tailored to the application under consideration.
Symmetry is a typically required property of the aggregation process:
the final value does not depend on the order in which the input values
are aggregated. Many aggregation operators satisfy this property,
but for example the weighted average is not a symmetric function.
The main application areas of aggregation operators include decision
making, pattern recognition, and machine learning. A crucial point
in the aggregation process is the type of values to be processed.
Depending on the values to be estimated by the experts, we can consider
three main types of input data to be aggregated: (\emph{i}) a value
in {[}0,1{]}, with higher numbers indicating higher potential risk
(crisp numbers); (b) an interval in {[}0,1{]}, in case the expert
cannot precisely quantify her opinion (interval). Linguistic expressions
modeled as different types of fuzzy sets also work in our platform:
for example high risk, low risk, medium risk, very high risk.

\section{Measuring systemic risk from expert opinions}

This section presents our approach to modeling systemic risk with
the use of expert opinions and aggregation operators. As was discussed
in the previous sections, the case of assessing systemic risk consists
most often of a set of interdependent characteristics. The relations
can depend for example on the type of risk or the market segment considered.
This ought to also impact how expert evaluations about risk are aggregated.
Herein, we present our model for measuring systemic risk in a problem
in which the system can be decomposed into a hierarchical form, in
which the decomposed characteristics are interdependent. We start
by describing the representation of this problem in terms of a Fuzzy
Cognitive Map (FCM), which aims at capturing the level of risk and
the interdependency among various characteristics. With an FCM as
its basis, we design the aggregation procedure to account for levels
and interdependencies through the use of the Choquet integral. The
outcome of the approach is an aggregation procedure allowing for the
measurement of systemic risk at various levels of the system, ranging
from continents to economies to market segments.

\subsection{Fuzzy Cognitive Maps }

As our main purpose is measuring/estimating systemic risk by utilizing
expert evaluations and opinions, we need to consider aggregation operators
from two perspectives: (\emph{i}) what are the different means to
aggregate data obtained from experts in different formats, and (\emph{ii})
how can we model the interrelated effects of different factors in
a system to obtain an overall risk estimation. To support the above
defined tasks in systemic risk measurement, we view the macro-financial
environment in terms of multiple market segments, which all exhibit
risk levels (i.e., risk identification) and interrelations to other
segments (i.e., risk assessment). Accordingly, we aim at describing
the system as a network, in which nodes represent risk severity for
different segments and edges among nodes represent the level of impact
of one component on the other. The values required to construct the
map is to be specified by experts by either a numerical value of a
linguistic expression (e.g., ``strong effect''). The final measure
of systemic risk is to be calculated by aggregating the information
from the different nodes through the interrelation values. Ideally,
we also aim at identifying an approach that not only provides a mathematical
model and a numerical risk estimation, but can also be used to offer
a visual representation of the different relationships within the
considered system. To fulfill these needs, we tap into the concept
of Fuzzy Cognitive Maps (FCMs).

In the seminal article, \citet{Kosko1986} defined FCMs with the aim
of representing complex systems in a structured way, which at the
same time allows for assessing the actual state, as well as tracking
and estimating possible changes in the system. At the general level,
a FCM is a directed graph with nodes as factors related to a general
concept and edges defining the relationships among nodes. The main
advantages of FCMs over traditional cognitive maps lies in the fact
that they can represent different degrees of influence of one factor
on another, and can be used to understand the potential consequences
of a change in the state of a node of the map (\citet{Montibeller2009}).
Consequently, one dominant line of applications of FCMs aims at modeling
changes taking place in dynamic and complex systems. A typical example
of this is a financial system comprised of a large number of interacting
entities. For example, in case of modeling systemic risk in a continent,
the system includes several subsystems (i.e., countries), which can
be seen as complex and dynamic systems on their own, with interrelated
components (i.e., market segments). In case of these dynamically changing
systems with highly interrelated components, we can face different
limitations by applying traditional modeling and analytical techniques. 

In many applications, FCMs are utilized for supervising complex processes
taking place in a system. Based on the obtained understanding of the
system, the overall goal many times is to translate it into a failure
detection or prediction system. Typical cases in the literature mainly
concern engineering systems or chemical processes taking place in
complex industrial systems (e.g., \citet{Mendoncca2013}). The common
factors among these applications are dynamic, nonlinear relationships
between system components. As was already pointed out, the modeling
of this type of systems requires methods that can utilize human experience
and the knowledge of domain experts (\citet{Aguilar2005}). In the
above engineering examples, experts with decades of experience can
develop a holistic view of the complex process taking place in the
system. The acquired knowledge makes them capable of identifying different
relationships between different components of a system that are hardly
recognizable with only collected historical data. Additionally, this
knowledge is not necessarily expressible in terms of, for example,
IF-THEN rules to form a basis of a classical expert-based decision
support system. In these cases, FCMs provide a framework to capture
experts' knowledge in a graphical representation that is more intuitive
(\emph{i}) for the experts to communicate their knowledge, and (\emph{ii})
for the decision maker to understand the behavior of the system as
a whole.

The application of FCMs in economics and finance is not widespread
yet, although we can identify a handful of contributions. In the context
of modeling stock market investments, \citet{Lee1997} developed an
new, bi-directional inference system based on FCMs with the main goal
of representing highly unstructured decision making problems. \citet{Koulouriotis2001}
attempt to model the complex system of agents affecting stock-market
behavior in national economies to provate organizations. The simulation
results show promising forecasting accuracy, although the authors
acknowledge a number of shortcomings of their model. \citet{Carvalho2004}
designed a qualitative, rule-based FCM approach to model dynamic economic
systems, specifically they use the example of modeling the behavior
of the European Central Bank with regard to interest rates.

\subsection{Mapping systemic risk from expert opinions}

The previous discussion can be formulated in a similar way when considering
financial systems with financial supervisors as the experts. The knowledge
of experts can be utilized in assessing the relationship between different
sectors of a financial system, specifically to what extent a given
sector affects another sector. The natural interpretation of a cognitive
map corresponds to that of a directed graph: nodes represent concepts
(i.e., components of a system) while edges represent causal relationships
between nodes. In the FCM approach, weights on the edges are incorporated
to represent the strength of the relationship between two components.
As our main goal is to estimate the level of systemic risk in a complex
system, the FCMs will be of a special type of weighted directed graphs,
as they will represent, on the general level, a hierarchical structure
of the components of the system. The top level of the hierarchy consists
of a single node with the associated value as the level of systemic
risk. On the second level of the hierarchy, the main components of
the system are included. For instance, in estimating the systemic
risk in Europe, this second level would consist of the individual
countries. At the next level, the sub-components or sectors of the
second-level components are included. In general, the set of components
can differ for different higher level nodes, such as including the
macroeconomy, banking sector, financial markets, etc. We can define
the required number of hierarchy levels in a similar way. In the following
formulation and later in the presented examples, we will discuss the
case of hierarchies with two levels additionally to the systemic risk
parent node for the sake of presentation; the formulation can easily
be extended to an arbitrary number of hierarchy levels.

The main component of our hierarchy, and correspondingly the central
node of the FCM, is the node $SR$ representing overall systemic risk.
The next level consists of the main components of the system, $S_{1},\ldots,S_{t}.$
Subsequently, every $S_{i}$ is associated with a set of sub components
$s_{i,1},\ldots,s_{i,p_{i}}$. The set of sub components is not necessarily
the same for every node on the higher level, so we formulate the problem
in a general way. Every node is associated to a corresponding value
of the actual level of risk in that specific system-component. Additionally,
the connections between the nodes represent the relationship among
different sectors of the underlying system. For example, if $S_{i}$
and $S_{j}$ are connected with edges with associated weights$w_{i,j}$and
$w_{j,i}$, then the higher the value of the edges, the larger the
extent is to which risk in the first sector results in increased risk
in the second sector. In the basic formulation, the weights of the
map are assumed to take a value in the $\left[0,1\right]$ interval
with $0$ indicating no effect and $1$ indicating maximum possible
effect (i.e., the two sectors are maximally interconnected). It is
important to note that $w_{i,j}$and $w_{j,i}$ are not equal; for
example, if we consider two countries in Europe, the effects on each
other are not necessarily symmetrical. According to these observations,
the most general representation of a FCM is an adjacency relation
matrix. which is not necessarily symmetrical, that defines the weights
on the edges with the actual values of estimated risk in different
components of the systems in the diagonal ($w_{i,i}$). 

\[
S=\left(\begin{array}{ccccc}
w(SR,SR) & w(SR,S_{1}) & ... & ... & w(SR,s_{t,p_{t}})\\
w(S_{1},SR) & w(S_{1},S_{1}) & ... & ... & ...\\
... & ... & ... & ... & ...\\
... & ... & ... & ... & w(s_{t,p_{t-1}},s_{t,p_{t}})\\
w(s_{t,p_{t}},SR) & ... & ... & w(s_{t,p_{t}},s_{t,p_{t-1}}) & w(s_{t,p_{t}},s_{t,p_{t}})
\end{array}\right)
\]

The process of employing FCMs for assessing systemic risk is performed
in the following steps that will be worked out in the subsequent parts
of this section:
\begin{enumerate}
\item The knowledge captured in the map is provided by experts (i.e., financial
supervisors). The experts are asked to fill in the elements of the
described adjacency matrix. This step can be performed regularly,
with the set of experts not necessarily being the same in every evaluation.
Naturally, the experts are not expected to provide estimations about
all the values in the matrix, but only of the sectors and dependencies
of the system they are confident about. As a result, we obtain a FCM
for every expert with associated confidence level for the provided
values. Based on these individual FCMs, we need to create an aggregated
FCM that incorporates the information by taking into consideration
their confidence values and the nature of the underlying hierarchy.
At time $t$, this step will result in the initial FCM, $S_{t}^{(0)}$.
\item Based on the FCM $S_{t}^{(0)}$ and the FCM $S_{t-1}$ created in
the previous evaluation process, we define $S_{t}$ as the actual
FCM representing systemic risk at time point $t$. The values of the
nodes will be updated by combining the information from both matrices
by using appropriate aggregation functions. In this case, appropriateness
refers to the need to aggregate interrelated quantities.
\item Using $S_{t}$, we can assess the risk in the system by making use
of traditional measures in directed graphs. This includes identifying
receiver and transmitter nodes in the map: components of the underlying
system that are (\emph{i}) highly vulnerable to risks taking place
in other connected components (receiver), and (\emph{ii}) components
that affect the risk in the system more significantly than others
(transmitter). We can also identify central nodes of the map by employing
different centrality measures. In our model, we can make use of the
special nature of the defined cognitive map, i.e. the underlying hierarchy
of the system. Using this, we can define global and local receivers,
transmitters and central nodes. To identify locally important nodes,
we can restrict the evaluation to a specific level of the hierarchy
to identify the component of the system that has the highest impact
on the risk on that level. Taking systemic risk in Europe as an example,
we can determine countries that contribute most to the risk level
of the system, and additionally analyze countries individually to
determine which sectors contribute most to the national financial
systems.
\item The nature of the FCM offers the possibility to create a graphical
representation of the dynamic evolution of the relationships among
the system components. We can employ visualization techniques to follow
the changes and identify the dependencies in the network that are
evolving in a way that points to a potential threat to the system
as a whole.
\end{enumerate}
Next, we consider the problem of creating the FCM at a fixed point
in time, so we avoid using an additional index for denoting the time
parameter. The organizational units are denoted by $o_{1},o_{2},\ldots,o_{k}$.
The experts associated to unit $o_{i}$ are denoted by $e_{i_{1}},...,e_{i_{k}}$.
Expert $e_{i_{j}}$provides the adjacency matrix $FS_{i_{j}}$. In
the literature, such as \citet{Khan2004}, it is usually assumed that
every expert has a single confidence level, $r_{i_{j}}$, that can
result in either (\emph{i}) the experts providing information only
about a small set of nodes and edges of the FCM about which they are
highly confident; or (\emph{ii}) the experts provide information on
a large set of nodes and edges but with low average confidence. While
the former situation enables formulating the final results of the
systemic risk estimation with fairly high confidence, yet with a small
sample, the latter case enables tapping into a larger sample, but
includes more uncertain opinions, and might hence hinder extracting
useful information. For this reason, a confidence value is used separately
for every piece of information (i.e., for each element of the adjacency
matrix of the FCM). For instance, expert $e_{i_{j}}$can state with
confidence $r_{i_{j}}(S_{l,}S_{m})$ that the effect of system component
$S_{l}$ on component $S_{m}$ is $w(S_{l},S_{m})$. The final value
will then be the confidence weighted average of the individual estimations:

\[
w(S_{l},S_{m})=\frac{\sum_{i,j}r_{i_{j}}(S_{l,}S_{m})w_{i_{j}}(S_{l,}S_{m})}{\sum_{i,j}r_{i_{j}}(S_{l,}S_{m})}.
\]

\subsection{Aggregating systemic risk from expert opinions}

This section draws upon the above described FCM setup as well as the
domain needs for systemic risk analysis. When modeling complex systems,
it is not reasonable to assume that the behavior of the components
of the system are independent from each other. In our case, we aim
at aggregating the risk evaluations in different components and combine
this evaluation with the interdependence of the components to obtain
values for the central nodes, i.e. the overall level of systemic risk.
There exists several methods to model the changes taking place in
the system through the impact of the nodes on each other on the existing
links. The most common method introduced already in the original paper
by \citet{Kosko1986} is to calculate the weighted average of the
values from nodes connected to the target node with a directed edge.
This simple aggregation procedure has several drawbacks, for example
it only allows for the representation of simple monotonic relationships
(\citet{Carvalho1999}), or the lack of temporal dimension in the
map (\citet{Zhong2008}). More importantly for our context, in case
a node is not assigned an initial value, it is estimated through the
weighted average without considering the interrelations between the
considered nodes. For this reason, from the broad literature on aggregation
operators, we use for systemic risk assessment an approach that enables
interdependencies and non-linear behavior to be captured when calculating
the initial value of a node in the fuzzy cognitive map. As we will
see later, the approach will not only provide a tool to incorporate
and model the complexity of the structure of a system in analyzing
the map but also allows for further analysis (by incorporating time
and different centrality values) to better understand the underlying
system and the spread of risk in it. To incorporate interaction of
different components in the aggregation process, a key feature of
an approach is to consider the effect of subsets of information rather
than individual elements. The basis for this approach is provided
by the construct of a fuzzy measure \citep{MurofushiSugeno1989}: 
\begin{defn}
A fuzzy measure $\mu$ on the finite set $N=\left\{ 1,2,...,n\right\} $
is a set function $\mu:P(N)\rightarrow\left[0,1\right]$ (where $P(N)$
is the power set of $N$) satisfying the following two conditions:\end{defn}
\begin{itemize}
\item $\mu(\phi)=0,\,\mu(N)=1$; 
\item $A\subseteq B$ implies that $\mu(A)\leq\mu(B)$.
\end{itemize}
The second condition enables representing in this framework measures
that do not satisfy the strong condition of additivity. In our context,
this means that we can model situations when the high value of a characteristic
of the financial system in itself does not indicate risk unless a
set of other characteristics show deviations from their usual values
at the same time. One of the most general formulations when using
monotone measures as the basis of aggregation can be described by
the Choquet integral \citep{Choquet1953,Marichal2002}. To formulate
the definition, we suppose that there are $n$ number of characteristic
measures $(c_{1},...,c_{n})$ based on which an evaluation is performed,
which results in the corresponding $(x_{1},...,x_{n})$ values. 
\begin{defn}
A discrete Choquet integral with respect to a monotone measure $\mu$
is defined as
\end{defn}
\[
C_{\mu}(x_{1},\text{\dots},x_{n})=\sum_{i=1}^{n}\left(x_{(i)}-x_{(i-1)}\right)\mu\left(C_{(i)}\right)
\]
where $x_{(i)}$ denotes a permutation of the $x_{i}$ values such
that $x_{(1)}\leq x_{(2)}\leq\ldots\leq x_{(n)}$ and $C_{(i)}=\left\{ c_{(i)},c_{(i+1)},\ldots,c_{(n)}\right\} $. 

Different well-known aggregation operators are special cases of the
Choquet integral, but also in its general form it starts to gain popularity
in different applications \citep{NarukawaTorra2007}. The basic properties
of the operator are determined by the monotone measure, such as symmetry,
additivity and linearity. The simplest case is when no interaction
is considered among the different evaluation criteria, which is reflected
in the monotone measure in a way that it takes only non-zero values
on singletons (i.e., individual criteria) and it is zero for all possible
combinations. This implies that the criteria are supposed to be independent
which is not necessarily a realistic assumption in our problem context,
although the aggregations mainly used in practice belong to this group
with the simple weighted average being the most prominent one. With
simple averages, we would look at the components of our system independently
and determine the importance of the components individually with respect
to overall systemic risk, whereafter an aggregated value over a threshold
would signal the existence of elevated levels of risk. Extreme measures
of this type are the maximum and minimum operators. Using the maximum
would imply that we consider a situation risky if there is at least
one component of the system that is elevated, while minimum is applicable
in a situation when we are concerned with the cases where all the
components exhibit elevated values. In general, different modifications
of the Choquet integral make it appropriate to integrate all the different
data types, including crisp numbers, intervals and linguistic expressions.
Although difficult to generalize, in some cases it is shown to be
the best possible aggregation tool \citep{LabreucheGabrisch2003}.
In the subsequent example cases, we will only use crisp numbers as
our main focus is on describing the conceptual framework, but it is
important to point out that it is straightforward to incorporate for
example fuzzy values in the aggregation as is done in \citep{TanChen2010}.

A more complex special case is the one when the pairwise interaction
between components of the model is considered additionally to the
individual effects. The Choquet integral in this case is 2-additive
and can be formulated as (see \citet{Grabisch1997}):

\begin{equation}
\begin{aligned}C_{\mu}(x_{1},\text{\dots},x_{n})= & \sum_{i=1}^{n}(v\left(c_{i}\right)-\frac{1}{2}\sum I(c_{i},c_{j}))x_{i}+\sum_{I(c_{i},c_{j})>0}I(c_{i},c_{j})\min(x_{i},x_{j})+\\
 & \sum_{I(c_{i},c_{j})<0}\mid I(c_{i},c_{j})\mid\max(x_{i},x_{j})
\end{aligned}
\label{eq:Choq}
\end{equation}

The interaction measure $I(c_{i},c_{j})$ can be defined in different
ways by transforming the measure for pairs into the $\left[-1,1\right]$
interval. A list of methods can be found for example in \citet{Marichal2000}.
If the interaction between pairs is zero, then they can be seen as
independent from each other, providing us the above described singleton
case. If the measure takes the value 1, it indicates that one component
strongly affect the other; intermediate values stand for different
degrees between no effect and strong effect. In our case, we will
restrict the interaction measure to the $\left[0,1\right]$ interval
with the maximum operator. In the following, we will describe how
Choquet integrals are used to aggregate values in the FCM.

As our main goal is to estimate the level of systemic risk, we will
start by describing the procedure of calculating $SR$ assuming that
all the other elements in the adjacency matrix are known. To make
use of the Choquet integral, first we need to identify the underlying
fuzzy measure. The domain of the defined measure is specified as all
the possible combinations of edges in the FCM but with only a small
subset of them associated to non-zero values. We will assign positive
value of the measure only to the set of paths, $P$, in the FCM that
has length smaller than or equal to 2 and the end-node of the path
is $SR$. This includes three types of paths:
\begin{itemize}
\item Edges in the form of $\left(S_{i},SR\right)$: connections between
first-level components and the final node
\item Edges in the form of $\left(S_{i},S_{j},SR\right)$: connections between
first-level components and the final node mediated by another first-level
component
\item Edges in the form of $\left(s_{j,p_{k}},S_{j},SR\right)$: connections
between second-level components and the final node mediated by a first-level
component
\end{itemize}
Based on these considerations, we can define a normalized fuzzy measure,
i.e. a fuzzy measure when the sum of the measure values is 1. In our
case, we will restrict the interaction measure to the $\left[0,1\right]$
interval, but it is important to note that the interpretation of our
case corresponds to the negative interactions in eq. \ref{eq:Choq}.
If the risk increases on any point in a path leading to the final
node, it will result in increased systemic risk. We do not require
the risk to be increased in every node on the path to achieve this.
Thus, the calculations are performed as follows:

\begin{equation}
C_{\mu}(x_{1},\text{\dots},x_{n})=\sum_{i=1}^{n}(v\left(c_{i}\right)-\frac{1}{2}\sum I(c_{i},c_{j}))x_{i}+\sum\mid I(c_{i},c_{j})\mid\max(x_{i},x_{j}),\label{eq:Choq-1}
\end{equation}

\hspace{-0.52cm}with $c_{i}$ as the edges with the final node as
the endpoint and the associated risk level $x_{i}$ in the starting
node. $I(c_{i},c_{j})$ denotes the overall effect on a path to the
final node and is calculated as the function of the values on the
two edges forming the path. There are several possibilities for combining
the two values reflecting what portion of risk we think is transferred
along the path. For example, if we use the product of the two values,
it will result in a lower rate of risk spread than using the maximum
of the two values. In the following, a numerical example is presented
to clarify the use of the Choquet integral in aggregating risk values
in the nodes of FCMs.
\begin{example}
To illustrate the use of the Choquet integral in aggregating the values
in a FCM, we look at a simple example of a system of two countries.
Fig. \ref{fig_example} depicts three different cases based on the
assumed relationship of the two countries. In the first map, there
is no relationship between the two countries, in this case the aggregation
with the Choquet integral is the same as the weighted average. Following
the previous discussion, we only have two elements in the set that
forms the basis of the fuzzy measure: the two edges of the FCM. In
this case, we obtain that 

\[
SR=\frac{0.3*0.5}{0.3+0.8}+\frac{0.8*0.3}{0.3+0.8}=0.35.
\]

In the second graph, we assume that there is a connection between
the two countries, but only one of the countries affects the other.
This model includes three paths in the set $P$: the two edges between
the countries ($p_{1}$ and $p_{2}$) and the final node and the path
from Country 2 to Country 1 to System ($p_{3}$). Additionally, we
have a non-zero measure value for the edge between the two countries
but it is not used directly in the systemic risk calculations. The
measure of the three paths is as follows: $\mu(p_{1})=\frac{0.3}{0.3+0.8+0.6*0.3},\mu(p_{2})=\frac{0.8}{0.3+0.8+0.6*0.3},\mu(p_{3})=\frac{0.3*0.6}{0.3+0.8+0.6*0.3}.$
The systemic risk is calculated based on eq. \ref{eq:Choq} as

$SR=\left(\frac{0.3+(0.6*0.3)/2}{0.3+0.8+0.6*0.3}-\frac{1}{2}\frac{0.3*0.6}{0.3+0.8+0.6*0.3}\right)0.5+\frac{0.8}{0.3+0.8+0.6*0.3}0.3+\frac{0.3*0.6}{0.3+0.8+0.6*0.3}\max(0.3,0.5)=0.38$.

We can see that the introduced relationship in the map increases the
value of systemic risk. In the third part of Fig. \ref{fig_example},
the map is further extended with a new connection between the two
countries, but with a lower associated value. An example of this type
of a situation is when there are two interconnected economies with
different size, such as the case of Germany (Country 2) and Poland
(Country 1). A risk in any of them will increase the risk in the other,
but as Germany is of greater importance, elevated risks in Germany
has more severs effects on Poland than the other way around (i.e.,
0.6 compared to 0.2). In this figure, we have four paths to be considered,
including the three from the previous example and $p_{4}$ as the
path from Country 1 to the final node through Country 2. They have
the following associated measure values: 
\[
\mu(p_{1})=\frac{0.3}{0.3+0.8+0.6*0.3+0.2*0.8},\mu(p_{2})=\frac{0.8}{0.3+0.8+0.6*0.3+0.2*0.8},
\]
\[
\mu(p_{3})=\frac{0.3*0.6}{0.3+0.8+0.6*0.3+0.2*0.8},\mu(p_{4})=\frac{0.2*0.8}{0.3+0.8+0.6*0.3+0.2*0.8}.
\]
 Hence, systemic risk is calculated as

\[
\begin{aligned}SR= & \left(\frac{0.3+(0.6*0.3)/2}{0.3+0.8+0.6*0.3+0.2*0.8}-\frac{1}{2}\frac{0.3*0.6}{0.3+0.8+0.6*0.3+0.2*0.8}\right)0.5+\\
 & \left(\frac{0.8}{0.3+0.8+0.6*0.3+0.2*0.8}-\frac{1}{2}\frac{0.2*0.8}{0.3+0.8+0.6*0.3+0.2*0.8}\right)0.3+\\
 & \frac{0.3*0.6}{0.3+0.8+0.6*0.3+0.2*0.8}\max(0.3,0.5)+\frac{0.2*0.8}{0.3+0.8+0.6*0.3+0.2*0.8}\max(0.3,0.5)=0.39
\end{aligned}
\]
which shows again an increase in the overall systemic risk value.

\begin{figure}[H]
\centering{}\includegraphics[scale=0.1]{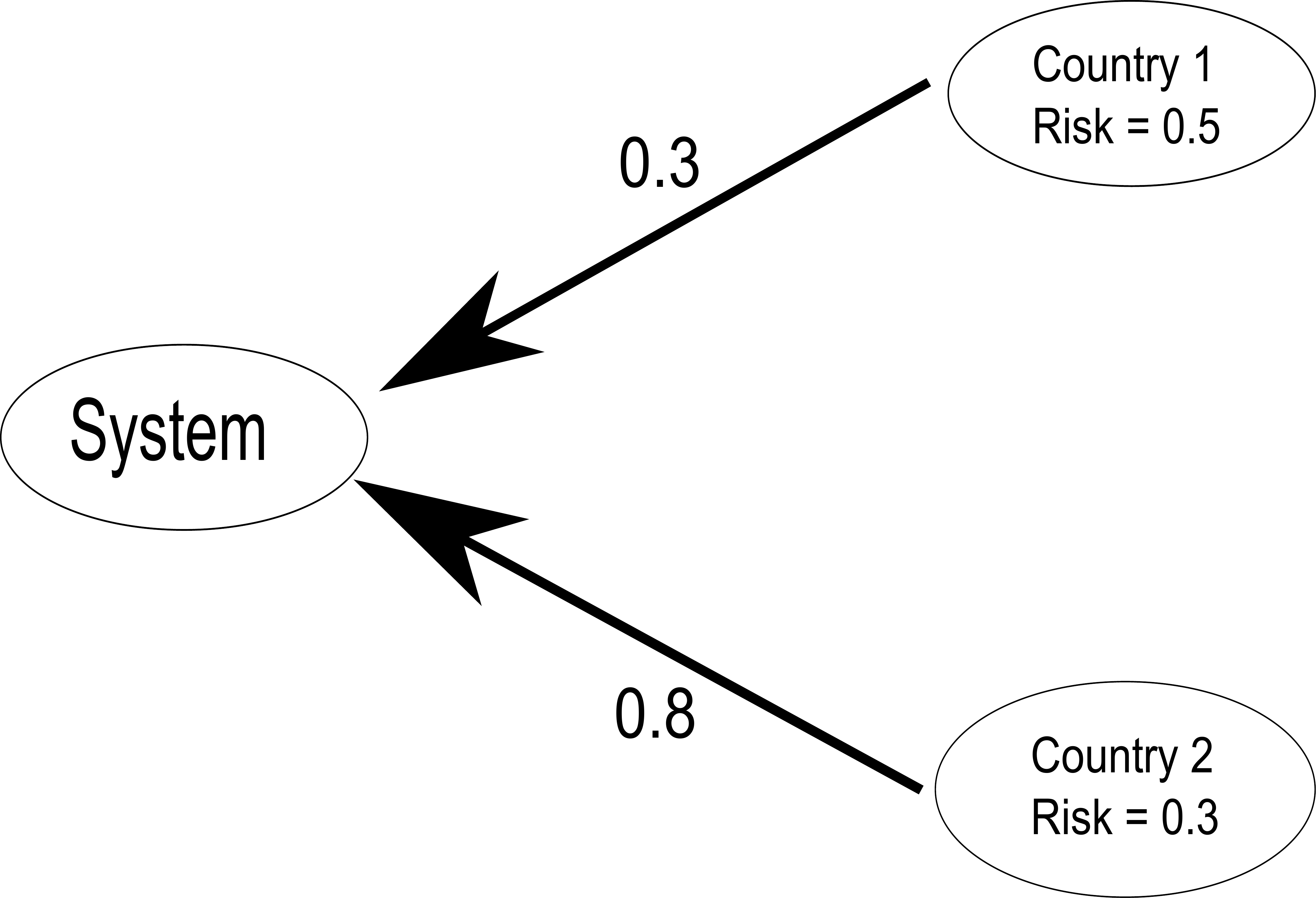}\includegraphics[scale=0.1]{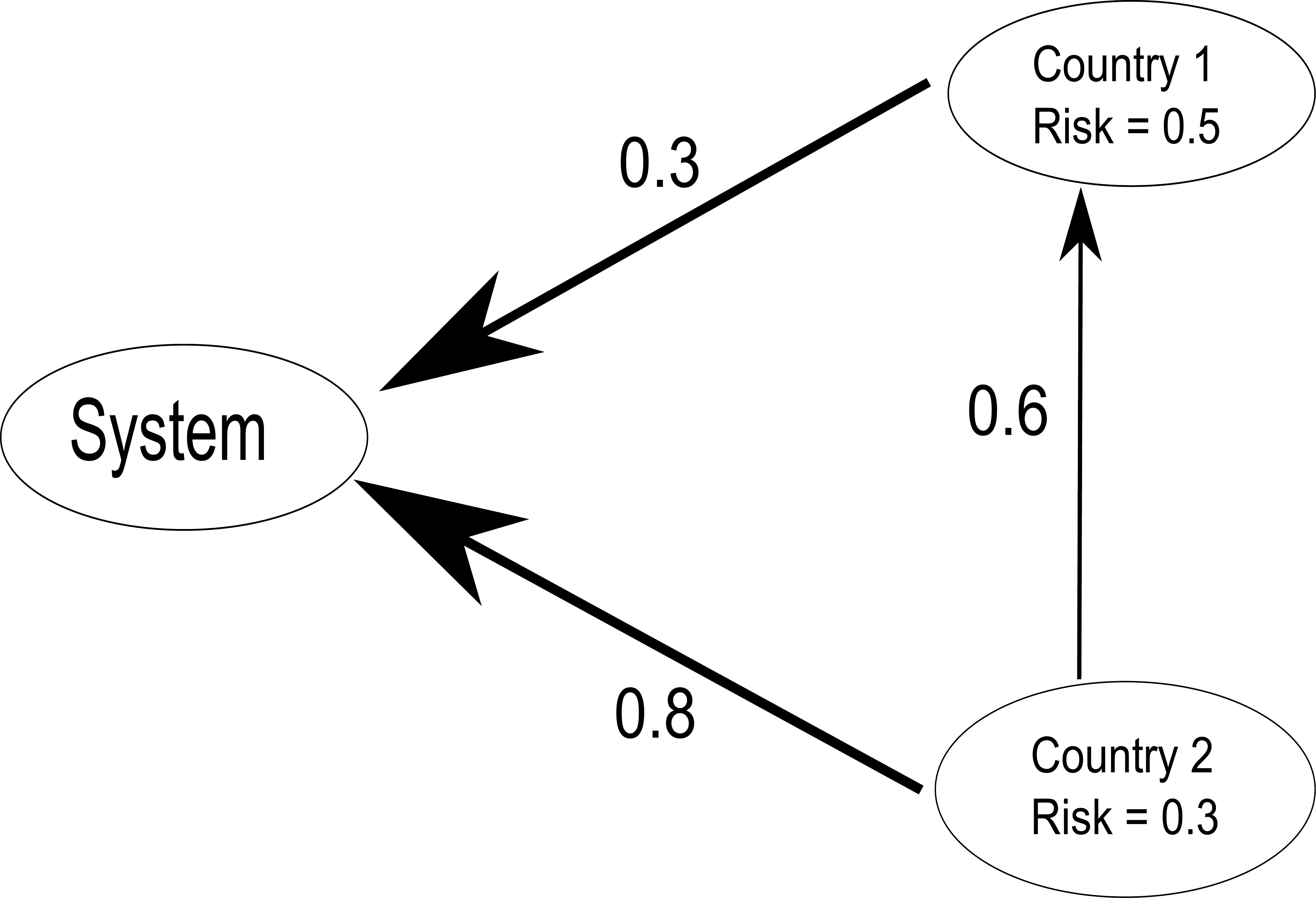}\includegraphics[scale=0.1]{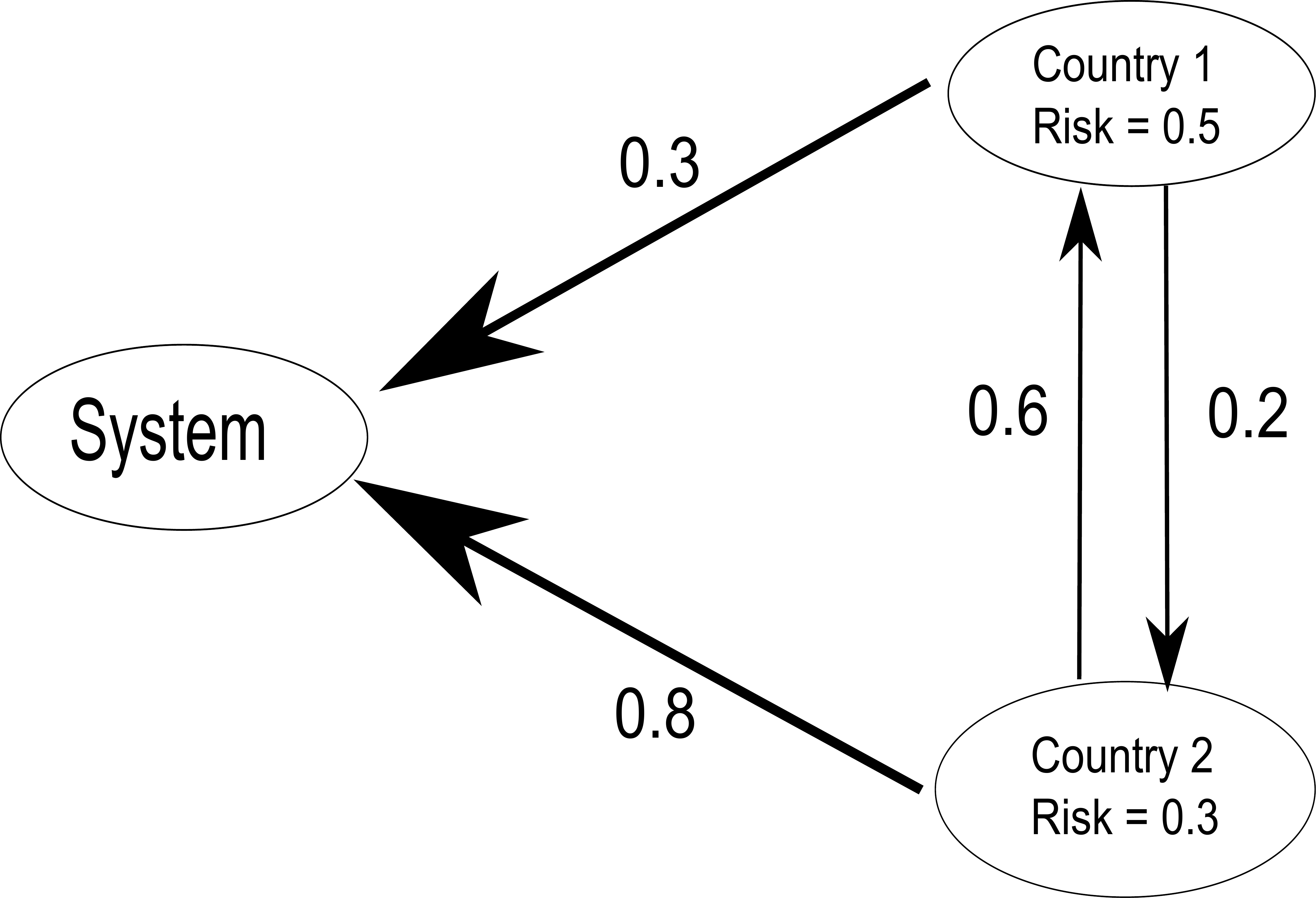}\protect\caption{An illustrative example of three types of interconnections.}
\label{fig_example}
\end{figure}

\end{example}
The described procedure illustrates the most straightforward application
of Choquet integral based aggregation in FCMs. The same thinking can
be applied to estimate the risk values at the first level of the system
hierarchy based on the connection to second-level sectors and the
risk levels in those sectors. This would result in a smaller number
of estimations to be provided by the experts in order to estimate
systemic risk. A fuzzy measure can be specified individually for every
node at the first level of the system hierarchy. This allows to calculate
the risk estimation by identifying the edges and paths consisting
of two edges with that specific node as the endpoint. In the next
step, systemic risk can be calculated based on the estimated risk
values in the nodes. 

A more important and novel way of utilizing Choquet integrals in the
fuzzy cognitive map is to consider the time dimension in the development
of the underlying system. In the above description, the underlying
assumption is that the risk present in any segment of the system contributes
at the same time to the systemic risk. In reality, we could look at
the problem in a way that takes into consideration the length of considered
paths in the aggregation. Considering the above example, we could
assume that the risk present in Country 2 directly contributes to
the present systemic risk, while it takes some time until this risk
affects the system as a whole, indirectly through Country 2's effect
on Country 1. In this respect, the aggregated systemic risk obtained
in the example evaluates a situation that will take place in a future
time-point; if we assume that the delay of spreading risk from one
node to a neighboring node is one time unit, then the example estimates
the development of the system in two time units from now. With the
similar reasoning, we could estimate (or ``forecast'') systemic
risk at different time points by considering paths with different
lengths in the construction of the fuzzy measure. Formally, to estimate
the state of the system $k$ time units from now based on the adjacency
matrix of the FCM at time point $t$, a sequence of fuzzy measures,
$\mu_{t}^{k}$ and corresponding Choquet integrals can be defined
by considering non-zero values in the measure paths in the FCM that
have length not greater than $k$ and has the final node ($SR$) of
the map as the endpoint of the path. Based on each measure we can
obtain an estimation of the risk values in different nodes of the
network at any point between now and $k$ time units later. As in
the case of paths of length 2, an even more important problem here
is to define the fuzzy measure to combine the value associated to
the edges on paths with different lengths. We can perform these operations
considering $\mu_{t}^{k}$ to be a $k$-additive monotone measures
described in \citep{Grabisch1997} and moving beyond the complexity
of the 2-additive Choquet integral as discussed above. For instance,
considering the product of values as the final path value for large
$k$'s in general results in low effect values meaning that after
a point we do not gain any new information by increasing the possible
length of considered paths. While using the maximum of the individual
values on the path would increase the systemic risk value significantly
after every step. It always depends on the understanding of the underlying
domain to determine for how many steps it is still meaningful to forecast
based on the present situation. In highly fluctuating and rapidly
changing systems the recommended value should be lower than in rather
stationary systems.

\subsection{Exploiting the mapping and its aggregation}

This section discusses practical challenges and considerations when
using FCMs and the Choquet integral in conjunction with expert evaluations.
First, we view requirements on the experts, in order to support consistent
and meaningful evaluations. Then, we discuss both quantitative and
qualitative analysis of the expert evaluations and their aggregations.

\paragraph{Requirements on the experts}

The approach presented in the previous section offers a relatively
straightforward procedure to combine several FCMs created by individual
experts into an aggregate form to represent the overall knowledge
of the experts. In a possible setting that can be envisioned as a
typical application, the approach can be used as the way of facilitating
knowledge sharing within organizations. The experts working in the
organizations, are working in different units with each unit focusing
on a specific topic. These topics (or a subset of them) form a system
that is depicted in the form of the FCM. Experts are asked to comment
on the strength of the relation between some components of the system.
This information reflects the individual domain knowledge of experts.
Consequently, it is not rare that the individual FCMs constructed
using the matrix provided by an expert cannot be directly used to
estimate systemic risk. As experts are most often knowledgeable only
in a limited number of components of a complex system, in order to
claim that they are confident enough regarding the estimations, the
provided matrices are expected to be sparse (or contain several values
with low confidence level). For instance, an expert in an organizational
unit with a focus on component $S_{i}$ of the system is likely to
provide confident estimations regarding the relation between this
component and the others, but not necessarily between two components
different from $S_{i}$. We can conclude that the involvement of multiple
experts in a systemic risk evaluation process of a complex system
is: (\emph{i}) sufficient in the sense that it can result in higher
reliability of the estimations and (\emph{ii}) necessary in the sense
that there is no single expert who can provide a credible assessment
of the whole system (although the individual evaluations can be used
independently to estimate the level of risk in components of the system).

The only requirement specified for the experts is to provide the estimations
in a way that forms a connected network. This is ensured by requiring
them to specify a value at a low level of the system hierarchy, and
that the associated node has to be connected to the main system node
through their antecedent node in the hierarchy. In practice, this
requires experts to connect each node for which the risk level is
evaluated to the system node highest in the hierarchy. As specified
above, experts are not required to estimate the risk in sectors that
themselves are assumed to be composed of a set of segments, although
they have to specify the relationship of this node to the other nodes.
This in turn makes it unnecessary to employ consistency measures for
assessing expert evaluations. For instance, if the experts are allowed,
as in the general description, to specify every element in the adjacency
matrix of the FCM, they can unintentionally define conflicting information,
which would create problems in specifying the fuzzy measure in a consistent
way. The perspective to expert opinions that we take incorporates
aspects of a general behavioral approach, but does still not directly
incorporate the communication and interaction of experts for convincing
each other. Yet, we do include feedback, as experts may receive the
aggregated assessment prior to their judgment, which can make them
rethink their individual evaluations and indirectly increase the consensus
without being of direct focus, as was shown in \citet{Van1974}. The
feedback can take the form of a simple report containing the final
FCM or a detailed report containing a comparison of the expert's evaluation
with other opinions.

\paragraph{Quantitative and qualitative analysis of the map}

The FCM representation allows for various types of quantitative analysis
of risk present in the underlying system. As the FCM can be seen as
a weighted, directed graph, we can utilize different measures from
graph theory. Besides traditional aggregation operators, we can make
use of the structure of the aggregation procedure introduced in the
FCM to identify the most important or central nodes of the system.
The first question to understand about the system may be the overall
level of interconnectedness in the FCM. For this purpose, the definition
of the density of a graph can be used: in general, if the number of
nodes of the map is $N$ and the number of edges is $E$, then the
density is $\nicefrac{E}{N\left(N-1\right)}$. By making use of the
special, hierarchical structure of our FCM, we can modify this measure
in two ways: (\emph{i}) decrease the value of the denominator as the
hierarchy puts restrictions on the pairs of nodes that can be connected;
and (\emph{ii}) make use of the fact that the FCM is a weighted graph
and use the weight of an edge in estimating the overall connectedness.

Two traditional measures for directed graphs have specific meaning
in the proposed FCM representation: the out-degree and in-degree of
a node. Additionally to the level of systemic risk, two important
issues need to be analyzed: (\emph{i}) which receiver nodes of the
FCM (i.e., components of the system) are most vulnerable to risk,
and (\emph{ii}) which transmitter node possess the highest threat
towards other nodes. The traditional approach to determine these important
nodes makes only use of the values of the edges in the map. The nodes
can be identified using the in-degree and out-degree of a node, $s$:
\[
\text{in}(s)=\sum_{t\in FCM,t\neq s,}w(t,s),\thinspace\text{out}(s)=\sum_{t\in FCM,t\neq s}w(s,t).
\]
The sum of these two quantities specifies the centrality of a node
in the map (i.e., degree): $c(s)=\text{in}(s)+\text{out}(s)$. There
are several ways to improve these measures. First, to assess the level
of risk propagation from a node, we can consider not only the direct
links from this node to others but also paths with length at least
two to estimate the potential effect of the node in its neighborhood.
In a similar manner, the evaluation of receiver status can incorporate
information on the indirect relationship from nodes that are not directly
connected through an edge but through a path with predefined length.
This analysis can make use of Choquet integral based aggregation in
a similar way as it was described to determine systemic risk. For
instance, taking the problem of vulnerability of a node, this concept
naturally translates to the final value of systemic risk for the final
node, as it is interpreted as how much risk the system receives from
its components. For any arbitrary node $s$, a fuzzy measure and the
corresponding Choquet integral, $C_{s,2}$, can be defined by assigning
non-zero weights to paths with length not more than two and having
$s$ as the endpoint. Additionally to the product function that is
used in the previous subsections to combine effects on a sequence
of edges, we can use for example different $T$-norms or triangular
norms (for a detailed overview see for example \citet{Grabisch2009})
that provide a general framework for the conjunction operation, including
the $\min$ operator or the product norm:

\begin{equation}
C_{s,2}^{T}(x_{1},\text{\dots},x_{n})=\sum_{t:(t,s)\in FCM}(v\left(t\right)-\frac{1}{2}\sum_{(q,t)\in FCM}T(t,q))w(t,t)+\sum\mid T(t,q)\mid\max(w(t,t),w(q,q)).\label{eq:choq}
\end{equation}
To incorporate higher levels of indirect effects between nodes, we
can use $C_{s,k}^{T}$, where paths ending in node $s$ and with length
at most $k$ are included. Assuming that the risk can be potentially
transferred through an edge in one time unit, this definition can
provide an estimation to what extent the node can be seen as a $k$-receiver,
i.e. what is the potential degree of vulnerability after $k$ time
periods.

Beyond quantitative network analysis, we also provide means for visualization.
As networks constitute an inherently complex source of information,
we ought to support in exploring the data through an overview and
and drill-down analysis, as well as overall communication. Drawing
upon the visual analytics paradigm, we aim at supporting analytical
thinking through interactive visualizations of the map and its information
products, where the operative term is interaction. Networks constitute
of nodes for entities and relationships among them in matrix form,
where an edge could be the link between nodes $S_{1}$ and $S_{2}$,
for instance. With matrices of directed and weighted graphs, we have
links $w(S_{1},S_{2})$ and $w(S_{2},S_{1})$ representing the size
of the link from $S_{1}$ to $S_{2}$ and from $S_{2}$ to $S_{1}$.
The matrix is of size $n^{2}$, where $n$ is the number of nodes
in the network. Hence, each node is described by its relationship
to each other entity, such as $S_{1}\in\mathbb{R}^{n}$. Yet, the
complexity of a graph is often decreased by constraining the existence
of links. To support qualitative analysis of networks, we follow \citet{Sarlin2013SWIFT}
in requiring two features: (\emph{i}) interactive and (\emph{ii})
analytical visualizations. 

Analytical visualization involves computational approaches for simplifying
data in a trustworthy manner. The notion of an analytical technique
for visualization differs from data analytics by rather using analytics
to reduce the complexity of data, with the ultimate aim of visualizing
underlying data structures. A commonly used dimension reduction approach
for networks is force-directed layouting, which positions nodes by
approximating node distances to their corresponding link strengths.
This seeks to uncover the structure of the network in terms of more
and less densely connected areas and their relation. Still, it has
a number of limitations and remedies, such as hairball visualizations
and cases of weighted networks with few strong but many weak connections.
The coupling of visual interfaces with interaction techniques goes
to the core of information visualization and visual analytics. Rather
than a final stop, a visual interface is a mere starting point for
data exploration and understanding, and requires hence the support
of means for interaction. This becomes evident in the visual information
seeking mantra by \citet{Shneiderman1996}: ''\emph{Overview first,
zoom and filter, then details-on-demand}''. The visualization provides
merely a high-level overview, and should hence be manipulated through
interaction to zoom in on a portion of items, eliminate uninteresting
items and obtain further information about requested items. Thus,
a large share of the revealed information descends from manipulating
the medium, which not only enables better data-driven communication
of information, but facilitates also tasks related to data exploration
and analysis. Despite the inherent value of visual interfaces to support
understanding and communication of data, the use of qualitative representations
is by no means a substitute to quantitative approaches to network
analysis.

\section{Two applications to macroprudential oversight}

This section illustrates the application of FCMs and aggregation operators
to two cases in macroprudential oversight in Europe. First, we describe
a pan-European application, which highlights systemic risk in various
countries and market segments. Second, we describe an application
within an individual country, which illustrates a more granular measurement
of risks.

\subsection{Measuring systemic risk in Europe}

\begin{figure}[H]
\begin{centering}
\includegraphics[width=1\columnwidth]{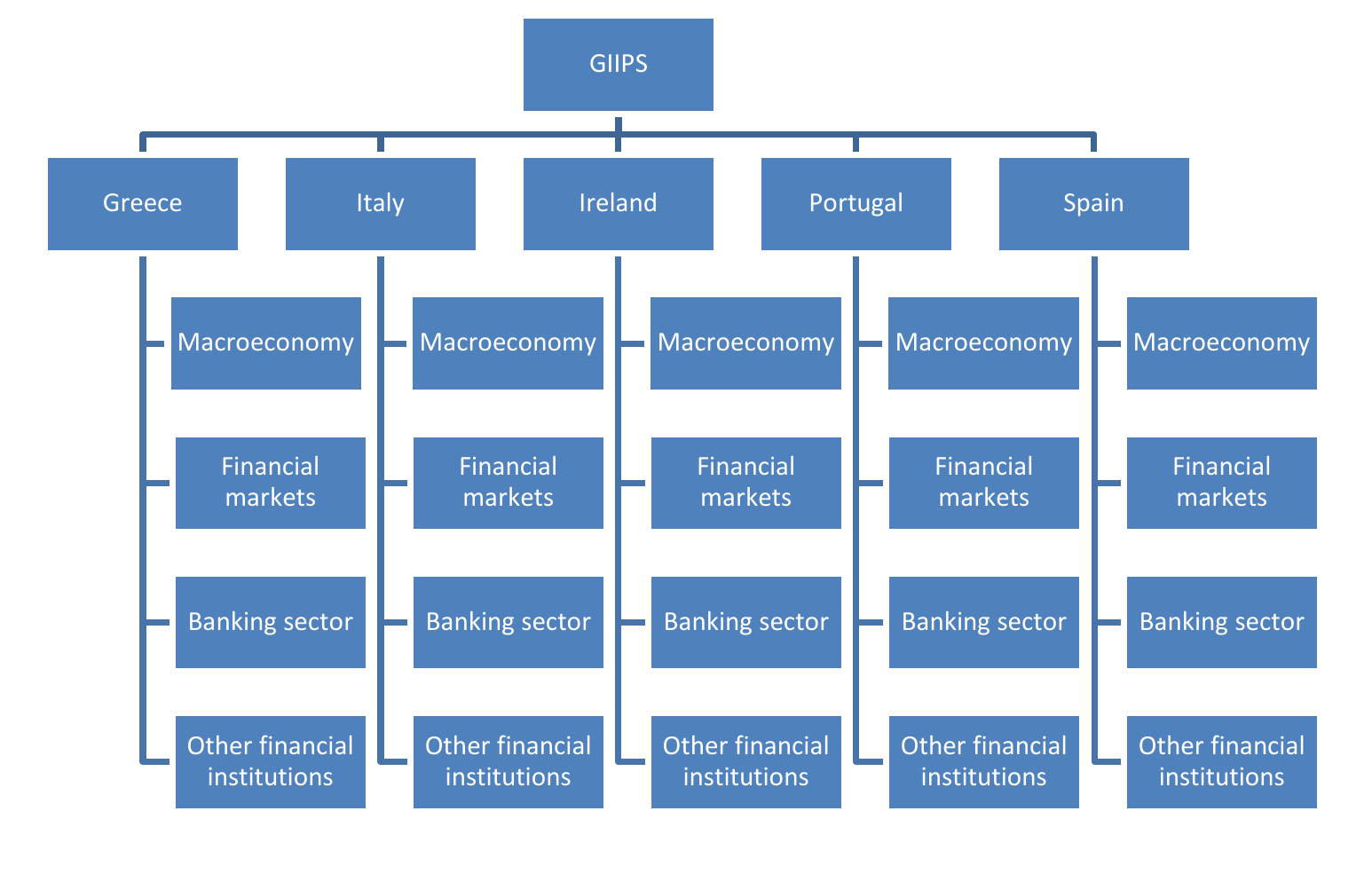}
\par\end{centering}

\centering{}\protect\caption{Decomposition of the systemic risk in Europe.}
\label{fig:europe_risk}
\end{figure}

Measuring systemic risk is a key concern of policymakers and supervisors
in Europe. Since the wake of the global financial crisis of 2007--2008,
the European institutional model for supervision and regulation has
adapted to the need for risk measurement and the implementation of
macroprudential policy. As can be exemplified by recently founded
supervisory bodies with the mandate of safeguarding financial stability,
a system-wide perspective to financial supervision is currently being
accepted and implemented as a common objective of governmental authorities
and supervisors, such as the European initiatives of the European
Systemic Risk Board (ESRB), the Single Supervisory Mechanism and the
Financial Stability Committee. Herein, we apply FCMs and aggregation
operators with a pan-European perspective to systemic risk. Figure
\ref{fig:europe_risk} describes the structure of the risk segments
that we assess. First, European systemic risk is decomposed to the
country-level, with five countries included: $S_{1},...,S_{5}$. Second,
for each country-specific segment $S_{i}$, we measure systemic risk
in four different sectors $s_{i_{1}},s_{i_{2}},s_{i_{3}},s_{i_{4}}$.
The sectoral risks include the macroeconomy, financial markets, banking
sector and other financial institutions. 

\begin{table}[H]
\protect\caption{Results from the hypothetical example.}

\begin{centering}
\begin{tabular}{ccccccc}
 & Greece & Italy & Ireland & Portugal & Spain & GIIPS\tabularnewline
\hline 
\hline 
Level of vulnerability & 0.73 & 0.58 & 0.61 & 0.71 & 0.77 & 0.72\tabularnewline
Out-degree & 9.42 & 7.88 & 5.04 & 8.5 & 9.86 & \tabularnewline
In-degree & 8.68 & 9.15 & 5.38 & 8.38 & 9.11 & \tabularnewline
Centrality & 18.10 & 17.03 & 10.42 & 16.88 & 18.97 & \tabularnewline
\end{tabular}\label{tbl:ex1}
\par\end{centering}

\end{table}

For a more comprehensible example, we focus on Greece, Italy, Ireland,
Portugal and Spain (henceforth the GIIPS). The evaluation process
starts with a set of experts expressing their insights on the level
of risk in the sectors of each of the five countries and the interdependencies
among sectors in individual countries. The experts are also asked
to evaluate the interrelations among countries, as well as impacts
from sectors to countries and countries to the GIIPS. We define the
risk of a sector as its vulnerability, or more precisely probability
of distress (i.e., not accounting for the impact of a shock), whereas
interrelations are defined as the impact of one segment on another
given the occurrence of a shock (i.e., not accounting for the probability
of a shock). In general, it would be possible to incorporate more
relationships (e.g., among sectors in different countries like the
impact of the Spanish banking sector on the Portuguese macroeconomy),
but in this example we restrict the input to the specified subset
of possible relationships. Moreover, in a more complex hierarchy (e.g.,
entire Europe), it is not reasonable to expect experts to be knowledgeable
or evaluate with a high confidence every country and sector. The example
illustrated herein provides thus a case of countries that oftentimes
naturally appear together in analysis related to banking and sovereign
risk, and consequently a high number of experts can be assumed to
have opinions on risks in all of the countries.

The outcome of this initial evaluation is the matrix of the information
provided by the experts and the matrix of corresponding confidence
values; at this point we obtain a matrix for every expert. The individual
evaluations (i.e., values associated with nodes and edges of the FCM)
can be aggregated according to the procedure described in the previous
section by taking the confidence weighted average for all the values
in the matrix. To obtain this matrix, we first aggregate the evaluation
of the experts for every cell using the confidence weighted average
of the completed evaluations. These values need to be normalized in
order to obtain the final adjacency matrix of the FCM. The following
phase is the actual assessment of the systemic risk level. Before
employing the Choquet integral, the confidence weighted average of
the evaluations is calculated as (for nodes $s$ and $t$):

\[
w(s,t)=\frac{1}{\sum_{e\in E}r_{i_{j}}^{e}}\sum_{e\in E}r_{i_{j}}^{e}w^{e}(s,t).
\]
Using these values, the final aggregated evaluation can be obtained
according to Eq. \ref{eq:choq}. The data that we utilized to illustrate
the use of the approach can be found in Table \ref{tbl:exdata1} in
the Appendix. In this table, the diagonal represents the estimated
vulnerability of a sector, while the other values reflect the impact
of one segment on another. For instance, the value 0.2 in the diagonal
implies that this sector is vulnerable only to a low extent, whereas
the value 0.8 shows significant vulnerability. The other values indicate
the strength of the interconnectedness among two segments. 

\begin{figure}[H]
\begin{centering}
\includegraphics[width=1\columnwidth]{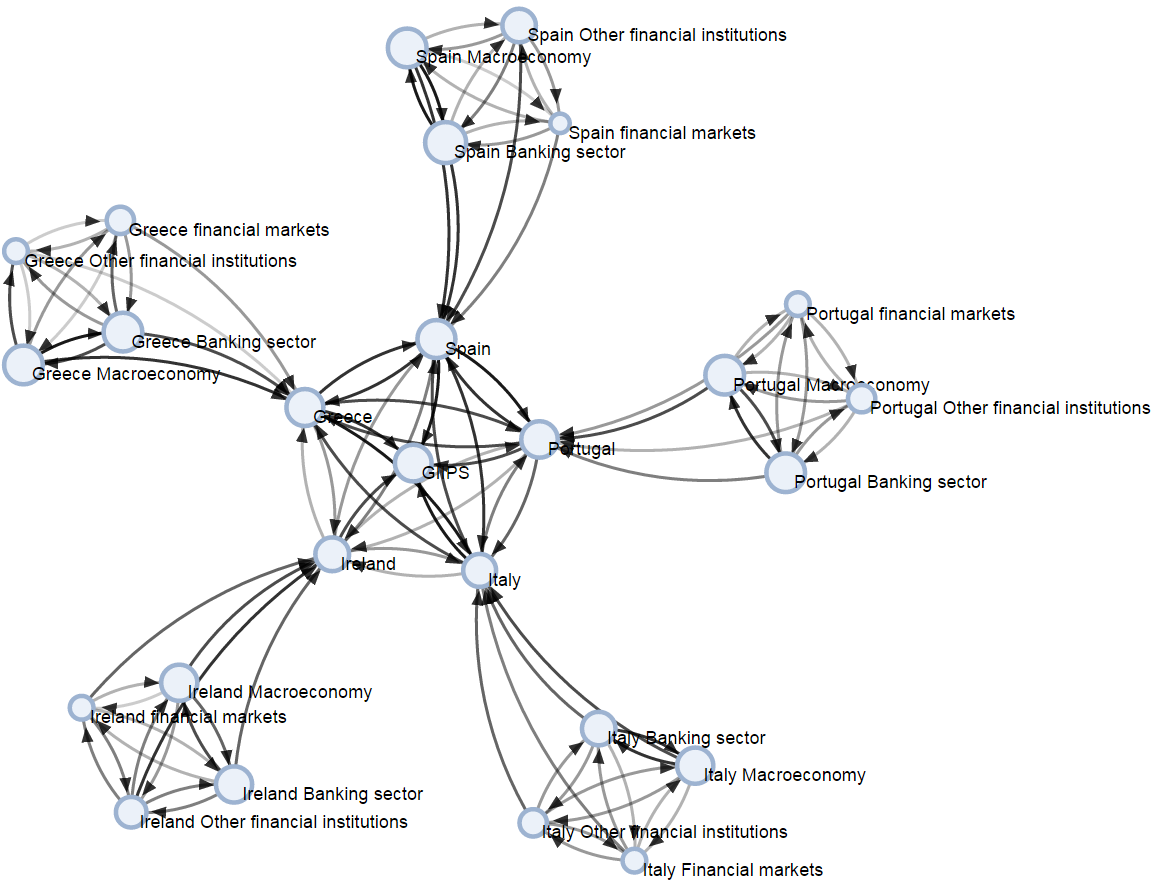}
\par\end{centering}

\centering{}\protect\caption{A network of systemic risk in Europe.}
\label{fig:europe_risk_network}
\end{figure}

The results of the hypothetical example can be found in Table \ref{tbl:ex1}.
As the numbers are not based on real data, although resembling reality,
it is worth noticing that the countries with high centrality, e.g.
Spain and Greece, have high levels of aggregated vulnerability. In
order to represent the results of expert evaluations and aggregations,
we make use of network visualization. Figure \ref{fig:europe_risk_network}
illustrates the network of countries and their sectors, as well as
the system level (i.e., GIIPS). The sectors, countries and GIIPS are
represented by nodes, whose size encodes level of vulnerability, whereas
their interrelations are shown through opacity of links. The figure
includes a screen capture from the publicly open web-based interactive
visualization platform VisRisk, in which this application has been
included as an instance.%
\footnote{The application is available here: http://vis.risklab.fi/\#/fuzzyAgg%
} This type of a visual interface could not only be an end product,
but also a means for supporting evaluations by experts and the feedback
of previous or aggregated opinions.

\begin{figure}[H]
\begin{centering}
\includegraphics[width=1\columnwidth]{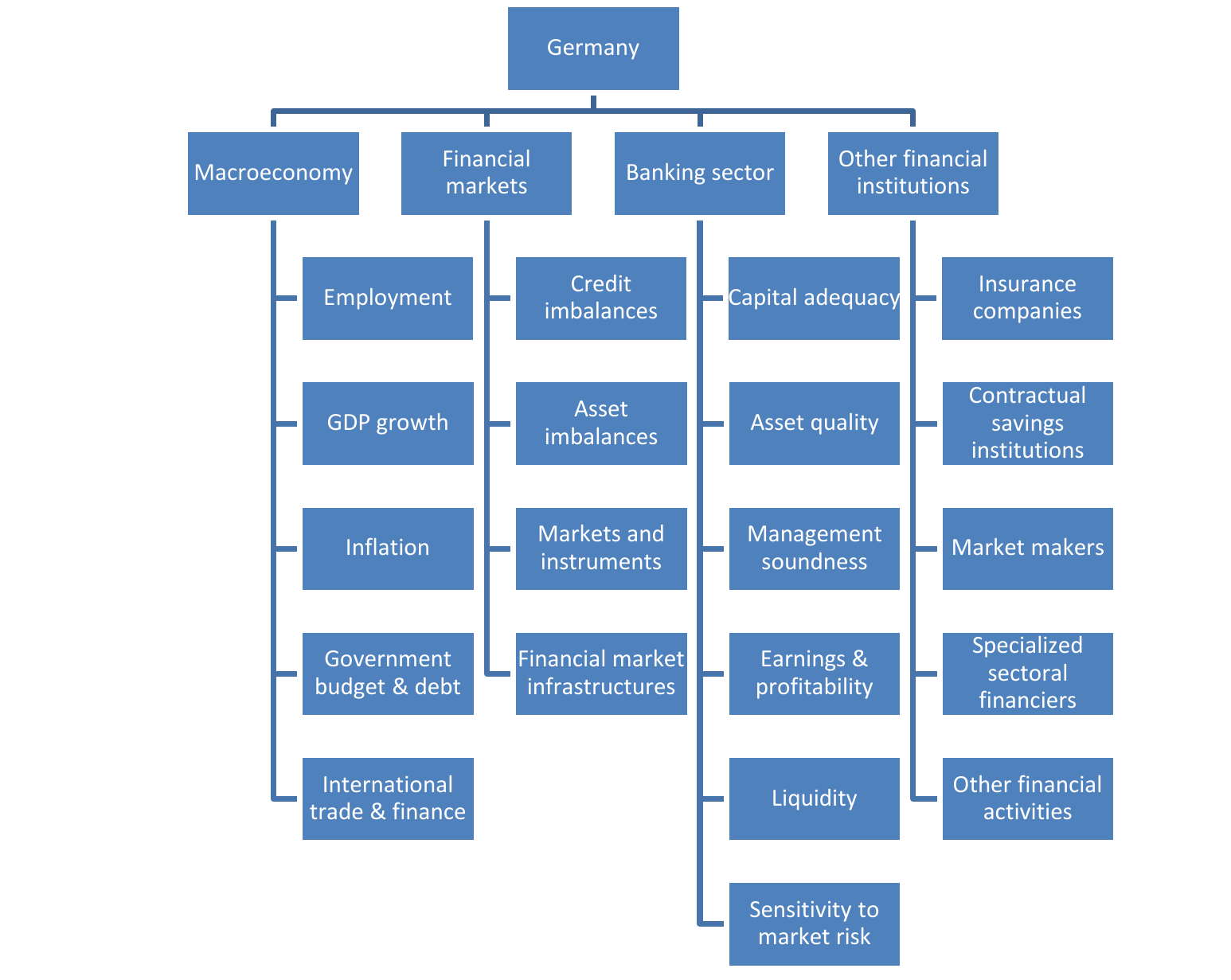}
\par\end{centering}

\centering{}\protect\caption{Decomposition of systemic risk in a country.}
\end{figure}

\subsection{Measuring systemic risk in a country}

The second application follows the previously discussed set-up, but
focuses on an individual country. Hence, we further breakdown the
risks to the macroeconomy, financial markets, banking sector and other
financial institutions into a number of sub-dimensions. In this application,
our segments $s_{1},\ldots,s_{t}$ refer to the risk in the sectors.
Accordingly, the second level of the hierarchy refers to sub-dimensions
pinpointing various types of risks in the sectors in question. Again,
to identify interconnectedness among various risks, and individual
sectors, experts are asked to represent interdependence on pairs of
components of the system. The process of finding the final systemic
risk value is the same as was described for the previous application.
The assessment of systemic risk in a country can be essential information
on its own, but at the same time it could be used as an input for
the first application. From a practical point of view, this would
put unreasonable requirement on the experts in terms of evaluations
to perform which can lead to unreliable results. To avoid this problem,
before the evaluation every expert can indicate the sectors and countries
she is competent to evaluate. Accordingly, the expert would be presented
only with the corresponding questions.

To illustrate the use of aggregation, we again provide a numerical
example based on the data provided in Table \ref{tbl:exdata2} in
the Appendix. As determining the aggregated importance values requires
only the use of a weighted average, we do not include details concerning
this step. We suppose that the values are already aggregated from
the individual evaluations obtained from experts, including their
confidence weights. Again, in Table \ref{tbl:exdata2} in the Appendix,
the diagonal represents the estimated vulnerability of a subdimension,
while the other values reflect the impact of one segment on another.
We consider a differing number of measures for every sector, but the
calculations are performed in a similar way according to the Choquet
integral formula. The final systemic risk value is 0.49, which indicates
a fairly low level of risk in this hypothetical example. 
\begin{table}[H]

\protect\caption{Results from the second example}

\begin{centering}
\begin{tabular}{cccccc}
 & Macroeconomy & FM & Banking sector & Other & Country\tabularnewline
\hline 
\hline 
Level of vulnerability & 0.29 & 0.62 & 0.43 & 0.47 & 0.49\tabularnewline
Out-degree & 6.21 & 5.35 & 5.17 & 4.94 & \tabularnewline
In-degree & 5.66 & 4.76 & 5.69 & 5.50 & \tabularnewline
Centrality & 11.87 & 10.11 & 10.86 & 10.44 & \tabularnewline
\end{tabular}\label{tbl:ex1-2}
\par\end{centering}

\end{table}

\section{Conclusion}

Multiple recent waves of systemic financial crises have sparked an
interest in measuring system-wide risks. The literature has predominantly
addressed the task by employing various statistical techniques using
market-based or accounting data. Yet, we reason that besides data-driven,
numerical approaches to quantifying systemic risk, one should also
tap into the knowledge of experts when assessing various dimensions
of vulnerability and risk. The experience of experts is a treasure
trove for revealing levels and trends of risks that are not necessarily
discoverable from available data sources, due to inter alia shadow
banking activities or other challenges in measurability of systemic
risk.

This paper puts forward a quantitative framework to incorporate the
knowledge of experts in the assessment of systemic risk. We propose
to model systemic risk as a network of risk segments through the Fuzzy
Cognitive Map (FCM), whose risks and interrelations are to be evaluated
by experts. The natural interpretation of the FCM corresponds to that
of a directed graph: nodes represent concepts (i.e., segments of a
system) while edges represent interrelations between nodes. Further,
we propose a procedure to aggregating risk to the highest level of
the system by incorporating levels of risk and interconnections among
the risk segments. The systematic aggregation approach to aggregate
expert evaluations of risk is based upon the Choquet integral. 

We have exemplified how the approach can aid macroprudential supervisory
bodies in making use of the information gathered from experts in their
organization, particularly for measuring systemic risk from expert
opinions in a European setting. First, we provided an estimation of
systemic risk in a pan-European set-up, where we systemic risk is
modeled at the European, country and sectoral level. Second, we also
illustrated the estimation of country-level risks, allowing for a
more granular decomposition. In the latter application, we model risk
at the level of one country, including its sectors and sub-dimensions
of the sectors. In conjunction with the applications, we also showed
both a quantitative and qualitative analysis of the systemic risk
measures by treating them as a network of nodes and edges. While the
former application uses standard measures of network importance to
describe nodes, the latter approach provides a visual interactive
interface for the systemic risk measures. The visualizations are available
as a complimentary web-based application.%
\footnote{The complimentary web-based applications are available here: http://vis.risklab.fi/\#/fuzzyAgg%
} 

The implications of this paper are twofold. First, the general introduction
of aggregation operators to systemic risk measurement sets a starting
point for the use of the rich, oftentimes tacit, knowledge in a policy
setting overall and their organizations in particular. We foresee
this to be an area of wide interest given the increased mandates and
overall responsibilities of financial supervisors, as well as the
discretionary nature of macroprudential policy. For instance, as discretionary
systems may be gamed yet rules struggle to properly account for risk,
\citet{AgurSharma2013} advocate a hybrid of discretion and rule-based
macro-prudential policy. Second, independent of the domain under analysis,
the theoretical contribution of the article provides means for a wide
range of applications combining FCMs and Choquet integrals to represent
and analyze complex systems of interrelated objects. In an increasingly
interlinked world, these types of systems are not a rare occurrence.
In light of the present paper, future research ought to tackle two
problems in need of further work: \emph{i}) the use and testing of
the proposed framework in practice, including all challenges involved
in collecting and using the expert knowledge for macroprudential purposes,
and \emph{ii}) the combination of the collection of expert knowledge
with interactive visual interfaces, in order to combine data-driven
analysis with expert judgment in a structured and collaborative manner.
\newpage{}

\section*{\textmd{\small{}\renewcommand\refname{References}}}

\section*{References}

\section*{\textmd{\small{}\bibliographystyle{plainnat}
\bibliography{References/references}
}}

\section*{\textmd{\small{}\newpage{}}}

\setcounter{table}{0}

\renewcommand{\thetable}{A.\arabic{table}}

\section*{Appendix: Data}

Table \ref{tbl:exdata1} and Table \ref{tbl:exdata2} provide the
data used in Section 4. In Table \ref{tbl:exdata1}, the abbreviations
stand for Macroeconomy (M), Financial markets (FM), Banking sector
(BS), and Other financial institutions (OFI). In Table \ref{tbl:exdata2},
the sectors are as follows:
\begin{itemize}
\item Macroeconomy: Employment (E), GDP growth (GDP), Inflation (I), Government
budget and debt (GBD), and International trade \& finance (ITF).
\item Financial markets: Credit imbalances (CI), Asset imbalances (AI),
Markets and instruments (MI), and Financial market infrastructures
(FMI).
\item Banking sector: Capital adequacy (CA), Asset quality (AQ), Management
soundness (MS), Earnings and profitability (EP), Liquidity (L), and
Sensitivity to market risk (SMR).
\item Other financial institutions: Insurance companies (IC), Contractual
savings institutions (CSI), Market makers (MM), Specialized sectoral
financiers (SSF), and Other financial activities (OFA).
\end{itemize}
\begin{table}[H]
\protect\caption{The data used in the first example in Section 4. }

~

\noindent \centering{}%
\begin{tabular}{ccccccc}
 &  & M & FM & BS & OFI & Greece\tabularnewline
\hline 
\hline 
\multirow{4}{*}{Greece} & M & 0.8 & 0.4 & 0.9 & 0.7 & 0.8\tabularnewline
 & FM & 0.2 & 0.4 & 0.4 & 0.3 & 0.4\tabularnewline
 & BS & 0.7 & 0.6 & 0.8 & 0.4 & 0.7\tabularnewline
 & OFI & 0.2 & 0.2 & 0.3 & 0.3 & 0.2\tabularnewline
\hline 
 &  & M & FM & BS & OFI & Italy\tabularnewline
\hline 
\hline 
\multirow{4}{*}{Italy} & M & 0.7 & 0.3 & 0.8 & 0.4 & 0.7\tabularnewline
 & FM & 0.3 & 0.3 & 0.4 & 0.4 & 0.5\tabularnewline
 & BS & 0.8 & 0.3 & 0.6 & 0.4 & 0.6\tabularnewline
 & OFI & 0.4 & 0.4 & 0.4 & 0.4 & 0.6\tabularnewline
\hline 
 &  & M & FM & BS & OFI & Ireland\tabularnewline
\hline 
\hline 
\multirow{4}{*}{Ireland} & M & 0.7 & 0.2 & 0.6 & 0.4 & 0.7\tabularnewline
 & FM & 0.3 & 0.3 & 0.3 & 0.5 & 0.6\tabularnewline
 & BS & 0.7 & 0.3 & 0.7 & 0.5 & 0.6\tabularnewline
 & OFI & 0.5 & 0.5 & 0.5 & 0.5 & 0.8\tabularnewline
\hline 
 &  & M & FM & BS & OFI & Portugal\tabularnewline
\hline 
\hline 
\multirow{4}{*}{Portugal} & M & 0.8 & 0.3 & 0.7 & 0.3 & 0.7\tabularnewline
 & FM & 0.3 & 0.3 & 0.4 & 0.3 & 0.4\tabularnewline
 & BS & 0.8 & 0.4 & 0.8 & 0.4 & 0.5\tabularnewline
 & OFI & 0.4 & 0.3 & 0.3 & 0.4 & 0.3\tabularnewline
\hline 
 &  & M & FM & BS & OFI & Spain\tabularnewline
\hline 
\hline 
\multirow{4}{*}{Spain} & M & 0.8 & 0.2 & 0.9 & 0.3 & 0.8\tabularnewline
 & FM & 0.3 & 0.2 & 0.4 & 0.3 & 0.5\tabularnewline
 & BS & 0.9 & 0.3 & 0.9 & 0.3 & 0.8\tabularnewline
 & OFI & 0.4 & 0.4 & 0.5 & 0.6 & 0.7\tabularnewline
\hline 
 & Greece & Italy & Ireland & Portugal & Spain & Europe\tabularnewline
\hline 
\hline 
Greece &  & 0.9 & 0.4 & 0.7 & 0.8 & 0.8\tabularnewline
Italy & 0.7 &  & 0.3 & 0.5 & 0.7 & 0.9\tabularnewline
Ireland & 0.3 & 0.4 &  & 0.3 & 0.4 & 0.7\tabularnewline
Portugal & 0.7 & 0.6 & 0.3 &  & 0.8 & 0.7\tabularnewline
Spain & 0.8 & 0.8 & 0.5 & 0.9 &  & 0.9\tabularnewline
\end{tabular}\label{tbl:exdata1}
\end{table}
\begin{table}[H]
\protect\caption{The data used in the second example in Section 4. }

~

\noindent \centering{}%
\begin{tabular}{cr@{\extracolsep{0pt}.}lccccccc}
 & \multicolumn{2}{c}{} & E & GDP & I & GBD & ITF &  & M\tabularnewline
\hline 
\hline 
\multirow{5}{*}{Macroeconomy (M)} & \multicolumn{2}{c}{E} & 0.3 & 0.8 & 0.4 & 0.8 & 0.4 &  & 0.3\tabularnewline
 & \multicolumn{2}{c}{GDP} & 0.8 & 0.3 & 0.5 & 0.8 & 0.5 &  & 0.4\tabularnewline
 & \multicolumn{2}{c}{I} & 0.4 & 0.4 & 0.3 & 0.5 & 0.5 &  & 0.3\tabularnewline
 & \multicolumn{2}{c}{GBD} & 0.6 & 0.6 & 0.4 & 0.2 & 0.4 &  & 0.5\tabularnewline
 & \multicolumn{2}{c}{ITF} & 0.7 & 0.7 & 0.6 & 0.7 & 0.3 &  & 0.4\tabularnewline
\hline 
 & \multicolumn{2}{c}{} & CI & AI & MI & FMI &  &  & FM\tabularnewline
\hline 
\hline 
\multirow{4}{*}{Financial markets (FM)} & \multicolumn{2}{c}{CI} & 0.3 & 0.8 & 0.5 & 0.3 &  &  & 0.6\tabularnewline
 & \multicolumn{2}{c}{AI} & 0.6 & 0.8 & 0.7 & 0.3 &  &  & 0.6\tabularnewline
 & \multicolumn{2}{c}{MI} & 0.3 & 0.4 & 0.5 & 0.5 &  &  & 0.3\tabularnewline
 & \multicolumn{2}{c}{FMI} & 0.3 & 0.3 & 0.5 & 0.4 &  &  & 0.2\tabularnewline
\hline 
 & \multicolumn{2}{c}{} & CA & AQ & MS & EP & L & SMR & BS\tabularnewline
\hline 
\hline 
\multirow{6}{*}{Banking sector (BS)} & \multicolumn{2}{c}{CA} & 0.5 & 0.6 & 0.7 & 0.5 & 0.5 & 0.4 & 0.6\tabularnewline
 & \multicolumn{2}{c}{AQ} & 0.5 & 0.4 & 0.6 & 0.7 & 0.7 & 0.7 & 0.5\tabularnewline
 & \multicolumn{2}{c}{MS} & 0.5 & 0.5 & 0.3 & 0.5 & 0.5 & 0.5 & 0.3\tabularnewline
 & \multicolumn{2}{c}{EP} & 0.7 & 0.7 & 0.5 & 0.4 & 0.6 & 0.5 & 0.4\tabularnewline
 & \multicolumn{2}{c}{L} & 0.8 & 0.6 & 0.5 & 0.4 & 0.4 & 0.5 & 0.5\tabularnewline
 & \multicolumn{2}{c}{SMR} & 0.7 & 0.7 & 0.5 & 0.6 & 0.7 & 0.4 & 0.4\tabularnewline
\hline 
 & \multicolumn{2}{c}{} & IC & CSI & MM & SSF & OFA &  & OFI\tabularnewline
\hline 
\hline 
\multirow{5}{*}{Other financial institutions (OFI)} & \multicolumn{2}{c}{IC} & 0.6 & 0.4 & 0.4 & 0.4 & 0.3 &  & 0.5\tabularnewline
 & \multicolumn{2}{c}{CSI} & 0.3 & 0.3 & 0.2 & 0.2 & 0.2 &  & 0.4\tabularnewline
 & \multicolumn{2}{c}{MM} & 0.3 & 0.2 & 0.4 & 0.2 & 0.2 &  & 0.3\tabularnewline
 & \multicolumn{2}{c}{SSF} & 0.2 & 0.2 & 0.2 & 0.5 & 0.2 &  & 0.3\tabularnewline
 & \multicolumn{2}{c}{OFA} & 0.2 & 0.2 & 0.1 & 0.1 & 0.2 &  & 0.3\tabularnewline
\hline 
 & \multicolumn{2}{c}{} & M & FM & BS & OFI &  &  & Country\tabularnewline
\hline 
\hline 
Macroeconomy (M) & \multicolumn{2}{c}{} &  & 0.8 & 0.7 & 0.7 &  &  & 0.4\tabularnewline
Financial markets (FM) & \multicolumn{2}{c}{} & 0.7 &  & 0.6 & 0.5 &  &  & 0.7\tabularnewline
Banking sector (BS) & \multicolumn{2}{c}{} & 0.7 & 0.4 &  & 0.7 &  &  & 0.6\tabularnewline
Other financial institutions (OFI) & \multicolumn{2}{c}{} & 0.6 & 0.4 & 0.7 &  &  &  & 0.6\tabularnewline
\end{tabular}\label{tbl:exdata2}
\end{table}

\end{document}